\documentclass[letterpaper,twocolumn,10pt]{article}
\usepackage{usenix}
\usepackage{epsfig}
\usepackage{amsmath}
\usepackage{amsfonts}
\usepackage{booktabs}
\usepackage{color}
\usepackage{nicefrac}
\usepackage[dvipsnames]{xcolor}
\usepackage{pifont}
\usepackage{multicol}
\usepackage{multirow}
\usepackage{xspace}
\usepackage{geometry}
\newgeometry{left=0.75in,right=0.75in,top=1in,bottom=1in}
\usepackage{graphicx}
\usepackage[font={small},tableposition=top]{caption}
\usepackage{subcaption}
\usepackage{enumitem}
\usepackage{microtype}
\usepackage{tablefootnote}
\usepackage{tikz}
\usepackage{pgf, import}
\usepackage{mathptmx}
\usepackage{url}
\usepackage{algorithm,algorithmic}
\usepackage[sort]{cite}
\usepackage[hidelinks]{hyperref}
\usepackage[all]{nowidow}

\usetikzlibrary{shapes.arrows}

\DeclareMathAlphabet{\mathcal}{OMS}{cmsy}{m}{n}

\newcommand{\sys}{Atom\xspace}
\newcommand{\Sys}{\sys}

\newcommand{\rulesep}{\unskip\ \textcolor{gray}{\vrule}\ }

\makeatletter

\let\c@table\c@figure
\makeatother

\newcommand{\com}[1]{}
\newcommand{\pubkey}{pk}
\newcommand{\seckey}{sk}
\newcommand{\enc}{\textsf{Enc}}
\newcommand{\encprf}{\textsf{EncProof}}
\newcommand{\dec}{\textsf{Dec}}
\newcommand{\keygen}{\textsf{KeyGen}}
\newcommand{\reenc}{\textsf{ReEnc}}
\newcommand{\shuffle}{\textsf{Shuffle}}
\newcommand{\reencprf}{\textsf{ReEncProof}}
\newcommand{\shufprf}{\textsf{ShufProof}}
\newcommand{\prf}{\pi}
\newcommand{\cost}{\mathcal{C}}

\newcommand{\msg}{m}

\newcommand{\ciphertext}{c}
\newcommand{\ciphertexts}{C}

\newcommand{\branch}{\beta}

\newcommand{\groupsize}{k}
\newcommand{\Z}{\mathbb{Z}}
\newcommand{\G}{\mathbb{G}}

\renewcommand{\paragraph}[1]{\smallskip\noindent\textbf{#1}}

\setlist[itemize]{itemsep=0pt, topsep=1pt,leftmargin=*}
\setlist[enumerate]{itemsep=0pt, topsep=1pt}

\begin{document}

\title{\sys: Horizontally Scaling Strong Anonymity}

\author{
  {\rm Albert Kwon}\\
  MIT
  \and
  {\rm Henry Corrigan-Gibbs}\\
  Stanford
  \and
  {\rm Srinivas Devadas}\\
  MIT
  \and
  {\rm Bryan Ford}\\
  EPFL
} 

\maketitle
\paragraph{Abstract}
\sys is an anonymous messaging system that protects against traffic-analysis attacks.
Unlike many prior systems,
each \sys server touches only a small fraction of the total
messages routed through the network.
As a result, the system's capacity scales near-linearly with the number of
servers.
At the same time, each \sys user benefits from ``best possible'' anonymity:
a user is anonymous among \textit{all} honest users of the system,
even against an active adversary who monitors the entire
network, a portion of the system's servers, and any number of malicious users.
The architectural ideas behind \sys have been known in theory,
but putting them into practice requires new techniques for
(1) avoiding heavy general-purpose multi-party
    computation protocols,
(2) defeating active attacks by malicious servers at
    minimal performance cost, and
(3) handling server failure and churn.

\sys is most suitable for sending a large number of short messages,
as in a microblogging application
or a high-security communication bootstrapping (``dialing'')
for private messaging systems.
We show that, on a heterogeneous network of 1,024 servers,
\sys can transit a million Tweet-length messages
in 28 minutes.
This is over $23\times$ faster than prior systems with similar
privacy guarantees.

\section{Introduction}\label{sec:intro}
In response to the widespread electronic surveillance of private communications~\cite{undersea},
many Internet users have turned to end-to-end
encrypted messaging applications,
such as Signal and OTR~\cite{otr}.
These encrypted messaging tools provide an effective way to
hide the {\em content} of users' communications from a network eavesdropper.
These systems do little, however, to protect users' {\em anonymity}.
In the context of whistleblowing~\cite{snowden,dissentv2},
anonymous microblogging~\cite{riposte}, or
anonymous surveys~\cite{anonize},
users want to protect their identities in addition to the
content of their communications.

Unfortunately, anonymity systems that protect against powerful global
adversaries typically cannot accommodate large numbers of users.
This is primarily due to the fact that traditional anonymity
systems only scale \emph{vertically}.
These systems consist of a handful of infrastructure servers that act collectively
as an anonymity provider;
the system can only scale by increasing the power of each participating server.
Systems based both on classical mix-nets~\cite{mixnet,vuvuzela,riffle} and
on DC-nets~\cite{dissentv2,riposte} suffer from this scalability challenge.

The Tor network~\cite{tor}, in contrast,
is an example of an anonymity system that scales \emph{horizontally}.
Tor consists of a network of volunteer relays, and
increasing the number of these relays increases
the overall capacity of the network.
This scalability property has enabled Tor to grow to handle
hundreds of thousands to millions of
users~\cite{tor_metric}.
However, the fact that Tor provides low-latency anonymity
also makes the system vulnerable to a variety of deanonymization
attacks~\cite{wang14-fingerprinting-defenses,ccs2013-usersrouted,
  wpes13-fingerprinting,ccs2012-fingerprinting,oakland2012-peekaboo,
  bauer:wpes2007,murdoch-pet2007}.

In this paper, we present \sys, an anonymous messaging system
that takes important steps
towards marrying the best aspects of these two architectural strategies.
Like Tor, \sys scales horizontally: adding more servers to
the network increases the system's overall capacity.
Like mix-net- and DC-net-based systems, \sys provides clear
security properties under precise assumptions.

\sys implements an \textit{anonymous broadcast primitive}
for short, latency-tolerant messages.
In doing so, \sys offers a strong notion of anonymity:
an adversary who monitors
the entire network, a constant fraction of servers, and any number of users
only has a negligible advantage at guessing which honest user sent which
message.

We target two applications in particular in this paper.
The first is an \emph{anonymous microblogging} application.
With \sys, users can broadcast short messages anonymously
to organize protests, whistleblow, or send other sensitive messages.
The second is a ``\emph{dialing}'' application:
many existing private messaging systems~\cite{vuvuzela, alpenhorn, pung}
require pairs of users to first establish shared secrets
using some out-of-band means.
\Sys can implement this sort of dialing system while providing
strictly stronger security properties than prior schemes can.

An \sys deployment consists of hundreds or thousands of
volunteer servers, organized into small groups.
To use the system, each user submits its
encrypted message to a randomly chosen entry group.
Once each server group has collected ciphertexts from a number of users,
the group shuffles its batch of ciphertexts,
and forwards a part of each batch to neighboring server groups.
After the servers repeat this shuffle-and-forward process
for a certain number of iterations, our analysis guarantees
that no coalition of adversarial servers can learn
which user submitted which ciphertext.
At this point, each group decrypts the ciphertexts
it holds to reveal the anonymized plaintext messages.

\sys's scalability comes from the fact that each
group of servers works {\em locally}, and only needs to handle
a small fraction of the total messages routed through the network.
In an \sys deployment routing $M$ messages using $N$ servers,
each \sys server processes a number of ciphertexts that grows as
$\tilde{O}(M/N)$.
In contrast, traditional verifiable-shuffle-based or
DC-net-based anonymity systems require each
server to do $\Omega(M^2)$ work,
irrespective of the number of servers in the system~\cite{dissentv2, riffle, riposte}.

The design of \sys required overcoming three technical hurdles.
First, in a conventional mix-net, each user produces an onion-style ciphertext,
in which her message is encrypted to each of the mix servers.
In \sys, the user does not know the set of servers its message will travel
through a priori, so she does not know which servers' keys to use
to encrypt her message.
Prior designs for distributed
mix systems~\cite{rackoff1993cryptographic,zamani2013towards,movahedi2014secure}
circumvented this problem with general-purpose multi-party
computation (MPC) protocols~\cite{ben1988completeness,goldreich1987play},
but these general methods are currently too inefficient to implement.
We instead use a new rerandomizable variant
of ElGamal~\cite{elgamal1984public} encryption, which allows groups of
servers in the network to collaboratively and securely decrypt and reencrypt
a batch of ciphertexts to a subsequent group.

Second, \sys must maintain its
security properties against {\em actively} malicious servers.
To protect against active attacks, we group the servers in such a way that
every group contains at least one honest server with overwhelming probability.
We then rely on the honest server to ensure that certain
invariants hold throughout the system's execution using two different
cryptographic techniques.
The first method relies on verifiable
shuffles~\cite{neff,groth-shuffle,crypto01-shuffle}, which can
proactively identify bad actors but is computationally expensive.
The second method is a novel ``trap''-based scheme, inspired by prior work
on robust mixing~\cite{khazaei2012mix}.
This scheme avoids using expensive verifiable shuffles,
but provides a slightly weaker notion of security:
a malicious server can remove $\kappa$ honest users from the system
(without deanonymizing them) with probability $2^{-\kappa}$,
thereby reducing the size of the remaining users' anonymity set.

Finally, \sys must handle network churn: with hundreds or
thousands of servers involved in the network, benign
server failures will be common.
Our mechanism allows us to pick a fault-tolerance parameter $h$,
and \sys can tolerate up to $h-1$ faults per group
while adding less than two seconds of overhead.
\sys also provides a mechanism for recovering from more than $h-1$ faults
with some additional overhead.

To evaluate \sys, we implemented an \sys prototype in Go,
and tested it on a network of 1,024 Amazon EC2 machines.
Our results show that \sys can support more than one million users sending
microblogging messages with 28 minutes of latency using 1,024 commodity machines.
(Processing this number of messages using Riposte~\cite{riposte},
a centralized anonymous microblogging system,
would take more than 11 hours.)
For a dialing application, \sys can support a million users
with 28 minutes of latency,
with stronger guarantees than prior systems~\cite{vuvuzela,alpenhorn}.

In this paper, we make the following contributions:
\begin{itemize}
  \item propose a horizontally scalable anonymity system
    that can also defend against powerful adversaries,
  \item design two defenses to protect this architecture
    against active attacks by malicious servers,
  \item design a fault-recovery mechanism for \sys, and
  \item implement an \sys prototype
    and evaluate it on a network of 1,024 commodity machines.
\end{itemize}

With this work, we take a significant step toward
bridging the gap between scalable anonymity
systems that suffer from traffic-analysis attacks, and centralized
anonymity systems that fail to scale.

\section{System overview}\label{sec:overview}
\sys operates by breaking the set of
servers into many small groups such that there exists
at least one honest server per group with high probability;
we call such group an \emph{anytrust}~\cite{dissentv2} group.
We then connect the groups using a carefully chosen link topology.

Communication in \sys proceeds in time epochs, or protocol \emph{rounds}.
At the start of each round, every participating user holds a plaintext message
that she wants to send anonymously.
To send a message through \sys, each user pads her message up to a fixed
length, encrypts the message,
and submits the ciphertext to a user-chosen \emph{entry group}.
Each entry group collects a predetermined number of user ciphertexts
before processing them.

After the collection, each group collectively shuffles
their set of messages.
The group collaboratively splits
the shuffled messages into several batches
and forwards each batch to a different subsequent group.
This shuffle-split-and-forward procedure continues for a number of iterations,
until a set of \textit{exit groups}
finally reveal the users' anonymized messages.
The exit servers can then forward these messages to either
a public bulletin board (e.g., for microblogging)
or an address specified in the message (e.g., for dialing).

Converting this high-level architecture into a working system design
requires solving a number of challenges:
\begin{enumerate}
  \item How does \sys provide protection against traffic-analysis attacks by
    a global adversary? (\S\ref{sec:network})
  \item To whom do clients encrypt their messages?
        If clients do not know which servers their messages
        will pass through, standard mix-net-style onion encryption
        will not suffice.
        (\S\ref{sec:basic})
  \item How does \sys protect users from malicious servers
    who deviate from the protocol? (\S\ref{sec:verifiable} and \S\ref{sec:trap})
  \item How does \sys remain resilient against server churn? (\S\ref{sec:server_churn})
\end{enumerate}

\subsection{Threat model and assumptions}
An \sys deployment consists of a distributed network of hundreds or thousands of servers,
controlled by different individuals and organizations.
A cryptographic public key defines the identity of each server,
and we assume that every participant in the system agrees
on the set of participating servers and their keys.
(A fault-tolerant
cluster of ``directory authorities'' could
maintain this list, as in the Tor network~\cite{tor}.)
Furthermore, we assume that the servers and the clients communicate over encrypted, authenticated,
and replay-protected channels (e.g., TLS).
A large number of users---on the order of millions---can participate
in each round of the \sys protocol.

We assume that the adversary monitors all traffic on the network,
controls a constant fraction $f$ of the servers,
and can control all but two of the users.
The adversarial servers may deviate from the
protocol in an arbitrary way, and
collude with each other and adversarial users.

\sys can provide availability in the presence of fail-stop server
faults (\S\ref{sec:server_churn}), but not against Byzantine server
faults~\cite{byzantine-generals-problem}.
Our fault-tolerance technique could mitigate
some availability attacks,
but we leave availability attacks out of scope.
That said, \sys \textit{does} protect users' anonymity in all cases.

Finally, \sys does not attempt to prevent intersection attacks~\cite{kedogan2002limits, danezis2004statistical}.
For example, if anonymous messages about a protest in Turkey
are only available when Alice is online,
then the adversary may be able to infer that Alice is sending those messages.
\sys does not protect against this attack, but
known techniques~\cite{wolinsky2013hang,hayes2016tasp}
can mitigate its effectiveness.

\subsection{System goals} \label{sec:model:goals}
\sys has three primary goals.

\paragraph{Correctness.}
At the end of a successful run of the \sys protocol
(i.e., a run that does not
abort), each server holds a subset of the plaintext messages sent through the
system.
Informally, we say that the system is {\em correct} if the union of the
message sets held at all honest servers contains every message
that the honest users sent through the system.

\paragraph{Anonymity.}
Following prior work~\cite{mixnet, kdd04-shuffle, dissentv2, riposte, riffle},
we say that \sys provides {\em anonymity} if an adversary who controls
the network, a constant fraction of the servers, 
and any number of users
cannot guess which honest user sent which message with
probability non-negligibly better than random guessing.
In particular, we require that the final permutation of
the honest users' messages
is indistinguishable from a random permutation.
Under this definition, a user is anonymous among \emph{all}
honest users, not just the users who share the same entry group,
even if there is only one honest user in an entry group.

\paragraph{Scalability.}
We say that an anonymity system is {\em scalable} if
the system can handle more users as the number of servers grows.
If there are $M$ messages and $N$ servers in the system,
we denote the number of ciphertexts each server processes
as $\cost(M,N)$.
We then require that holding $\cost(M,N)$ fixed,
$M$ scales linearly with $N$.

\subsection{Cryptographic primitives} \label{sec:crypto}
Atom relies on the following two cryptographic primitives.
We describe them at a high level here,
and give the details in Appendix~\ref{app:enc}.

\paragraph{Rerandomizable encryption.}
\sys uses a rerandomizable CPA-secure encryption scheme, which
consists of the following algorithms:
\begin{itemize}
  \item $(\seckey, \pubkey) \gets \keygen()$.
    Generate a fresh keypair.
  \item $\ciphertext \gets \enc(\pubkey, \msg)$.
    Encrypt message $\msg$ using public key~$\pubkey$.

  \item $\msg \gets \dec(\seckey, \ciphertext)$.
    Decrypt ciphertext $\ciphertext$ using secret key~$\seckey$.

  \item $\ciphertexts' \gets \shuffle(\pubkey, C)$.
    Rerandomize a vector $\ciphertexts$ of ciphertexts
    using public key $\pubkey$, and randomly permute the components of the vector.

  \item $\ciphertext' = \reenc(\seckey, \pubkey, \ciphertext)$.
    Strip a layer of encryption off of ciphertext $\ciphertext$
    using secret key $\seckey$ and add a layer of encryption
    using public key $\pubkey$.
    When $\pubkey = \bot$, this operation is the same as
    $\dec(\seckey, \ciphertext)$.
\end{itemize}
For \sys, we require an additional property that
the cryptosystem allow ``out-of-order'' decryption and reencryption.
That is, given an onion-encrypted ciphertext,
a server can decrypt one of the middle layers of encryption
and reencrypt the ciphertext to a different public key.

\paragraph{Non-interactive zero-knowledge proofs of knowledge (NIZKs).}
We make use of three NIZK constructions.
We use non-malleable NIZKs,
meaning that the adversary cannot use a NIZK for an input
to generate a different NIZK for a related input.
\begin{itemize}
  \item $(\ciphertext, \prf) \gets \encprf(\pubkey, \msg)$.
    Compute $\ciphertext \gets \enc(\pubkey, \msg)$ and generate
    a NIZK proof of knowledge of the plaintext $m$ corresponding to $\ciphertext$.
  \item $(\ciphertext, \prf) \gets \reencprf(\seckey, \pubkey, \msg)$.
    Compute \\ $\ciphertext \gets \reenc(\seckey, \pubkey, \msg)$
    and generate a NIZK proof $\prf$ that this operation was done correctly
    (cf.~\cite{chaum-pedersen}).
  \item $(\ciphertexts', \prf) \gets \shufprf(\pubkey, \ciphertexts)$.
    Permute the ciphertext set $\ciphertexts$ (using $\shuffle$) and
    generate a NIZK $\prf$ that $\ciphertexts$
    is a permuted version of $\ciphertexts'$, reblinded using $\pubkey$
    (cf.~\cite{neff,groth-shuffle,bayer}).
\end{itemize}
When the operations other than $\shuffle$ and $\shufprf$
are applied to a vector of ciphertexts $\ciphertexts$,
we apply the operation to each component of the vector.

\section{Random permutation networks} \label{sec:network}
In this section, we describe our solution to the first challenge:
how does \sys provide protection against traffic-analysis attacks by
a global adversary?
We use a special network topology that can provide
anonymity against an adversary
that can view the entire network and control any number of system users
but that \textit{controls no servers}.
Then, in \S\ref{sec:protocol}, we describe how to
protect against malicious servers.

\begin{figure}
  \centering
  \includegraphics[width=0.45\textwidth]{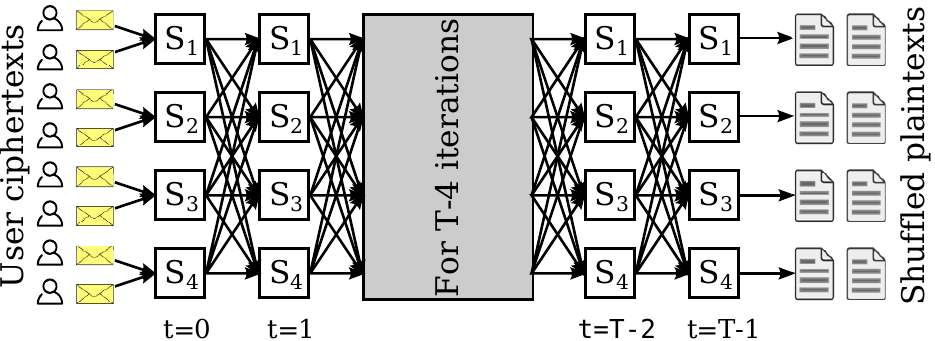}
  \caption{An \sys deployment with four servers ($S_1$ through $S_4$), arranged
          in the square topology. Each user submits her ciphertext
          to an entry server of her choice.
          The servers perform $T$ iterations of mixing before outputting the
          plaintexts.}
  \label{fig:square}
\end{figure}

We organize the \sys servers into a layered graph,
and each \sys server is connected to $\branch$ other servers
in the next layer,
where $\branch$ is a fixed branching factor.
An example topology is shown in Figure~\ref{fig:square} for $\branch = 4$.
In each protocol run, each user first chooses an \sys server
(an ``entry server'') and encrypts her message with the public key of the server
using a rerandomizable encryption scheme.
Each user then sends her ciphertext to the entry server.
Along with her ciphertext, the user also provides a NIZK to prove she
knows the underlying plaintext, to prevent a malicious user
from submitting a rerandomized copy of an honest user's ciphertext;
this would result in duplicate plaintexts at the end of the network,
which would immediately reveal the sender of the message.
We include the id of the server in the NIZK generation to prevent
a malicious user from resubmitting the exact copy of a ciphertext
and its NIZK to a different server.
The details of this NIZK are shown in Appendix~\ref{app:enc}.

We assume that each entry server receives the same
number of messages.
To achieve this in practice, an untrusted load-balancing server could direct
users to different entry servers.
We argue later in this section that every honest user of the system
is anonymous amongst \textit{all} honest users of the system.
Thus, a user's choice of entry group is not important for security.
Moreover, the untrusted server just directs the users to a server,
and does not actually route any messages.
It therefore cannot selectively remove users' messages
in an attempt to launch an intersection attack.

Once the \sys entry servers collect a certain pre-specified number of
ciphertexts, the servers then begin a mixing process that repeats for $T$ iterations,
where $T$ is a parameter chosen later on.
In each iteration of the mixing process, each server performs the following steps:
\begin{itemize}
  \item Randomly permute the set of ciphertexts.
  \item Divide the ciphertexts into $\branch$ batches of equal size.
  \item Reencrypt batch $i \in [1, \branch]$
    for the $i^{\text{th}}$ neighboring server
    and forward the reencrypted batch to that server.
\end{itemize}
The servers repeat this permute-and-reencrypt
process using the incoming batches. 
After $T$ iterations, each server in the network
holds a set of ciphertexts that has passed through $T$ other servers in
the network.
The \textit{exit servers} can then decrypt these ciphertexts and publish the corresponding
plaintexts.

\paragraph{Security analysis.}
We use a particular network
topology called a {\em random permutation network}~\cite{czumaj2014,czumaj2015,haastad2006square,czumaj1999delayed,czumaj2001switching}
to connect the servers.
When the servers are organized into such a network,
the output of the network after carrying out the protocol above
is a near-uniform random permutation of all input messages.
The adversary will thus have negligible advantage in
guessing which honest user sent which message over random guessing.
\sys is compatible with any good random permutation network.
In this work, we leverage prior analyses
to identify two simple candidate topologies:

\noindent\textbf{\textit{Square network:}}
H{\aa}stad studied the problem of permuting a square matrix of $M$ elements by repeatedly
permuting the rows and columns~\cite{haastad2006square}.
His analysis gives rise to a
random permutation network on $\sqrt{M}$ nodes in which each vertex shuffles $\sqrt{M}$ ciphertexts
and connects to $\sqrt{M}$ vertices on the subsequent layer (Figure~\ref{fig:square}).
H{\aa}stad demonstrated that this network produces a near-uniform random permutation after
only $T \in O(1)$ iterations of mixing.
In general, when we have many more messages than servers (in particular when $N < \sqrt{M}$),
we can have each server ``be responsible'' for multiple vertices in the network.
For example, $S_1$ and $S_2$ in $t = 0$ and $1$ in Figure~\ref{fig:square}
could be handled by a single server.

\noindent\textbf{\textit{Iterated-butterfly network:}}
It was shown that $O(\log M)$ repetitions
of a standard butterfly network yields an almost-ideal
random permutation network on $M$ elements~\cite{czumaj2014},
where each node in the butterfly network handles $O(1)$ messages.
We say ``almost ideal'' because the network produces
a random permutation on a constant fraction of the $M$ elements.
Adding a small constant fraction of dummy messages
to the system lets us use this network as
if it produced a truly random permutation~\cite{czumaj2014}.
Since a butterfly network has depth $O(\log M)$, the total depth of
the network is $O(\log^2 M)$.
If there are fewer than $M$ servers,
then we again have a single server emulate multiple nodes.
The resulting $N$-server network topology would have $O(\log^2 N)$ depth,
meaning there would be $T \in O(\log^2 N)$ iterations of mixing.

\paragraph{Efficiency.}
With $M$ total messages in the network, each server in each mixing iteration
handles $M/N$ messages.
For the square network, the number of mixing iterations satisfies ${T \in O(1)}$,
and thus each server only handles $O(M/N)$ ciphertexts total in the square network.
In the case of the iterated-butterfly network,
the number of messages each server handles is
${O(\frac{M}{N} \cdot \log^2 N)}$,
since there are ${T \in O(\log^2 N)}$ iterations.
For the rest of this paper, we focus on the square network,
since it will perform better in practice due to the shallower depth.
Both networks meet the scalability requirement
as both can handle more messages as we increase the number of servers
with a fixed $C(M,N)$.

\section{\sys protocol} \label{sec:protocol}
In this section, we extend the protocol of \S\ref{sec:network}
to defend against actively malicious servers
(i.e., servers who can tamper with the messages).
At a high level, we divide the servers into anytrust groups,
and replace the servers in the network with the groups.
Each group then simulates an honest server using our protocol,
and we provide mechanisms to detect malicious servers deviating from the protocol.

\subsection{Anytrust group formation} \label{sec:gen_anytrust}
The \sys servers are organized into groups.
As we will see later in this section,
the security of \sys relies on each group having at least one honest server
in this setting.
We ensure this by using a public unbiased
randomness source~\cite{bonneau2015bitcoin, sytascalable}
to generate groups consisting of randomly sampled servers.
We set the group size large enough to ensure that the probability
that all servers in any group are malicious is negligible.
Here, we make an assumption that the adversary controls at most
a particular constant fraction $f$ of the servers in the network.

For example, if we assume that the adversary controls
$f = 20\%$ of the servers, we can bound the probability that
any group consists of all malicious servers (``is bad'') by
choosing the group size $k$ large enough.
We compute:
$$\Pr[\text{One group of $k$ servers is bad}] \leq f^\groupsize,$$
and then use the union bound to compute
$$\Pr[\text{Any of $G$ groups of $k$ servers is bad}] \leq G \cdot f^\groupsize.$$

If we allow failure with probability at most
$2^{-64}$ with ${G = 1024}$ groups,
then we choose the group size $\groupsize = 32$ such that
${f^\groupsize \cdot G < 2^{-64}}$.
Finally, we replace the servers with the groups in the permutation
network.

New groups are formed at the beginning of each round.
In practice, this operation will happen in the background,
as the rest of the protocol is carried out.

\subsection{Basic \sys protocol} \label{sec:basic}
\begin{algorithm}[tb]
  \vspace{-3mm}
  \caption{Basic \sys group protocol}
  \label{alg:basic}
  \begin{flushleft}
  Form anytrust groups using the protocol in \S\ref{sec:gen_anytrust}.
  A group then takes a set of ciphertexts $\ciphertexts$ as input,
  either from the users if this server is in the first layer of the network
  or the groups in the prior layer, and executes the following.
  \end{flushleft}
  \begin{enumerate}
  \item \textbf{Shuffle}: \label{step:basic_shuffle}
    Each server $s$
    computes ${\ciphertexts' \gets \shuffle(\ciphertexts)}$ in order,
    and sends $\ciphertexts'$ to the next server.
  \item \textbf{Divide}:
    If $s$ is the last server of the group,
    then it divides the permuted ciphertext set
    $\ciphertexts'$ into $\branch$ evenly sized batches
    $(B_1, \dots, B_\beta)$,
    where $\branch$ is the number of neighboring groups.
    It sends $(B_1, \dots, B_\beta)$ to the first server of the current group.
  \item \textbf{Decrypt and Reencrypt}: \label{step:basic_dec}
    Server $s$ receives batches $(B_1, \dots, B_\beta)$
    from the previous server.
    In order, each server $s$
    computes $B'_i = \reenc(\seckey_s, \pubkey_i, B_i)$ for each batch
    where $\seckey_s$ is the secret key of the current server
    and $\pubkey_i$ is the group public key of the $i$th neighboring group.
    If there are no neighbors (i.e., this is the last iteration of mixing),
    then $\pubkey_i = \bot$ for all $i$.

    If $s$ is the last server of the group,
    then it sends all $B'_i$ to the first server in the $i^{\text{th}}$
    neighboring group.
    Otherwise, it sends them to the next server.
  \end{enumerate}
  \begin{flushleft}
  If $s$ is the last server of a group in the last layer,
  then $(B'_1, \dots, B'_\beta)$ contains the plaintext messages.
  \end{flushleft}
  \vspace{-3mm}
\end{algorithm}

We first describe our protocol that enables each group to collectively
shuffle and reencrypt the message, without protection against active attacks.
To send message $\msg$, the user first picks her entry group.
She then computes \\${(\ciphertext, \prf) \gets \encprf(\pubkey, \msg)}$,
where $\pubkey$ is the \emph{group public key};
this produces an onion-encrypted ciphertext encrypted using
the public keys of all servers in the group. For ElGamal, for example,
$\pubkey$ would be the product of the public keys of all servers in the group.
She then sends $(\ciphertext, \prf)$ to all servers in the entry group,
and $\prf$ is verified by all servers.

Once enough ciphertexts and proofs are received,
each group performs a specialized multi-party protocol
for shuffling the messages described in Algorithm~\ref{alg:basic}.
The security of this protocol relies on
Step~\ref{step:basic_shuffle} and Step~\ref{step:basic_dec}.
After Step~\ref{step:basic_shuffle},
the messages are permuted using
the composition of all servers' permutations.
Since the honest server's permutation is random and unknown,
the final permutation is secure as well.

Step~\ref{step:basic_dec} solves our second challenge:
to whom do the users encrypt their messages?
Here, we use the special out-of-order reencryption
property of our cryptosystem (\S\ref{sec:crypto} and Appendix~\ref{app:enc}).
A user only needs to encrypt for her entry group
(and need not know the path her message will take a priori),
and the entry group can reencrypt for the appropriate next group.
In particular, the first server decrypts a layer of encryption of a ciphertext
and reencrypts it for the next group of servers.
When the second server receives the resulting ciphertext,
it can decrypt its layer of encryption,
despite the fact that the ciphertext was last encrypted
with a different public key.
Since each message is simultaneously decrypted
and reencrypted for another group,
all messages remain encrypted under at least one honest server's key
until the last layer.
Thus, the adversary does not learn anything
by observing the traffic in intermediate mixing iterations.

\subsection{\sys with NIZKs} \label{sec:verifiable}

We now describe the two different mechanisms
that address our third challenge:
protecting against actively malicious servers.
First, each user generates her ciphertext and the corresponding
NIZK using $\encprf$, and submits them to all servers in their entry groups.
All servers then verify the NIZKs,
and report the verification result to all other servers in the group.
Our protocol then uses
{\em verifiable shuffles}~\cite{neff,groth-shuffle,bayer}
and {\em verifiable decryption}~\cite{chaum-pedersen}:
After each operation in Algorithm~\ref{alg:basic},
the server who shuffled or reencrypted the messages
proves the correctness of its operation using NIZKs
to all other servers in the group.
The honest server in each group will detect any deviation from the protocol.
Algorithm~\ref{alg:nizk} describes the details.

\begin{algorithm}[h]
  \vspace{-3mm}
  \caption{\sys group protocol with NIZKs}
  \label{alg:nizk}
  \begin{flushleft}
  A group receives a set of ciphertexts $\ciphertexts$ as input,
  as well as the proof verification results of the previous layer.
  \end{flushleft}
  \vspace{-4mm}
  \begin{enumerate}
  \item \textbf{Shuffle}:
    In order, each server $s$ does the following:
  \vspace{-1mm}
    \begin{enumerate}
      \item Compute
        ${(\ciphertexts', \prf) \gets \shufprf(\pubkey, \ciphertexts)}$,
        where $\pubkey$ is the current group public key.
      \item Send $(\ciphertexts', \prf)$ to all servers in the group.
        All servers in the group verify the proof $\prf$,
        and send the result of the verification to all servers
        in the group.
        All servers then check that every server in the group correctly
        verified the proof,
        and abort the protocol if any server reports failure.
    \end{enumerate}
  \item \textbf{Divide}:
    If $s$ is the last server in the group,
    then server $s$ divides the permuted ciphertext set
    $\ciphertexts'$ into $\branch$ evenly sized batches
    $(B_1, \dots, B_\beta)$,
    where $\branch$ is the number of neighboring groups.
    Server $s$ sends $(B_1, \dots, B_\beta)$ to the first server.
  \item \textbf{Decrypt and Reencrypt}:
    In order, each server $s$ does the following
    after it receives batches $(B_1, \dots, B_\beta)$ from the previous server:
  \vspace{-1mm}
    \begin{enumerate}
    \item
        Compute $(B'_i, \prf_i) = \reencprf(\seckey_s, \pubkey_i, B_i)$
        for each batch $B_i$,
        where $\seckey_s$ is the secret key of the current server
        and $\pubkey_i$ is the group public key of the $i^{\text{th}}$
        neighboring group.
      \item
        Send $\{(B'_i, \prf_i)\}_{i\in[\beta]}$
        to all servers in the current group.
        Send them to all servers in all the neighboring groups as well,
        if $s$ is the last server of the group.
        All servers that received the proofs $\{\prf_i\}_{i\in[\beta]}$
        verify them,
        and send the results of the verification to all servers in the
        current group, and the neighboring groups if $s$ is the last server
        in the group.
        Then, all servers who received the proofs check that
        all other servers who verified the proofs reported success,
        and abort the protocol if any server reports failure.
    \end{enumerate}
  \end{enumerate}
  \vspace{-5mm}
  \begin{flushleft}
  If $s$ is the last server of a group in the last layer,
  then $(B'_1, \dots, B'_\beta)$ contains the traps and the inner ciphertexts.
  \end{flushleft}
  \vspace{-3mm}
\end{algorithm}

\subsection{\sys with trap messages} \label{sec:trap}

Generating and verifying the zero-knowledge proofs imposes
a substantial computational cost on the servers.
The verifiable shuffle proposed by Neff~\cite{neff},
for instance, requires each server to perform
a number of exponentiations per element being shuffled.
Since every server needs to produce and verify a number
of these NIZK proofs, the extra computation can be burdensome.
Instead, we use a novel \emph{trap message}-based protection,
inspired by prior work~\cite{khazaei2012mix}.
In this variant, each user submits a ``trap'' ciphertext
with the ciphertext of her message.
If a server misbehaves, it risks tampering with a trap.
We then provide a distributed mechanism
to detect if a trap has been modified.

The malicious servers may get lucky and tamper with a real user message.
As such, the trap variant of \sys provides
a slightly weaker notion of security than the NIZK variant:
the adversary could remove (but not deanonymize) up to
$\kappa$ honest messages from the anonymity set with probability $2^{-\kappa}$.
If the number of honest users is large, as we expect in a
scalable system, this weaker anonymity property is almost as
good as the traditional anonymity property.
The NIZK variant of \sys can provide the stronger anonymity if needed.

This trap variant of \sys makes use of an extra anytrust group of servers, which
we call the \emph{trustees}.
The trustees first collectively generate a per-round public key for the group,
with each trustee holding a share of the matching secret key.
Users then encrypt their messages
using a double-enveloping technique~\cite{golle2002}:
each user first encrypts her messages using the trustees' public key
with an IND-CCA2 secure encryption scheme~\cite{naor1990public, rackoff1991non},
which ensures that the resulting ciphertext cannot be modified in anyway.
Then, the user encrypts the resulting ciphertext,
which we call \emph{inner ciphertexts},
with the group key of her entry group.
More precisely, to send a message $\msg$, a user
\begin{enumerate}
  \item encrypts $\msg$ using the trustees' public key $\pubkey_T$:\\
    ${\ciphertext_M \gets \texttt{"}\enc_{\textsf{CCA2}}(\pubkey_T, \msg) \| \texttt{M}}\texttt{"}$,
    where $\texttt{M}$ indicates that $\ciphertext_M$ is an inner ciphertext,
  \item picks an entry group. Let $gid$ be the index of the entry group,
  \item chooses a random nonce $R$ and
    generates a ``trap message'' as
    ${\ciphertext_T \gets \texttt{"}gid \| R \| \texttt{T} \texttt{"}}$,
    where $\texttt{T}$ indicates that $\ciphertext_T$ is a trap message,
  \item computes
    ${(\ciphertext_0, \pi_0) \gets \encprf(\pubkey, \ciphertext_M)}$,
    \\${(\ciphertext_1, \pi_1) \gets \encprf(\pubkey, \ciphertext_T)}$,
    and the cryptographic commitment $C_T$ of $\ciphertext_T$,
    where $\pubkey$ is the public key of the entry group
    (since the nonces are high-entropy,
    we can use a cryptographic hash like SHA-3~\cite{sha3} as a commitment),
  \item sends $(\ciphertext_0, \pi_0)$ and $(\ciphertext_1, \pi_1)$
    in a random order along with $C_T$ to all servers in her entry group.
\end{enumerate}

The servers verify $\pi_0$ and $\pi_1$ when they receive them.
Once a group collects enough ciphertexts and commitments,
each group carries out Algorithm~\ref{alg:basic},
treating each ciphertext (real or trap) as an independent message.
At the end of the protocol, the last server checks
its subset of messages. It then forwards
(1) each trap message to all servers in the group indicated by the $gid$ field,
and (2) each inner ciphertext to all servers in the group
chosen by a deterministic function
that will load-balance the number of ciphertexts forwarded to a group
(e.g., using universal hashing).

Each server reports the following to the trustees:
\begin{enumerate}
  \item a bit indicating whether
    every trap commitment has a matching trap and vice-versa.
  \item a bit indicating that
    all inner ciphertext has been forwarded correctly
    (e.g., hash is the expected value),
    and that there are no duplicate inner ciphertexts.
  \item the number of traps and inner ciphertexts.
\end{enumerate}

\begin{figure}[tb]
  \centering
  \includegraphics[width=0.49\textwidth]{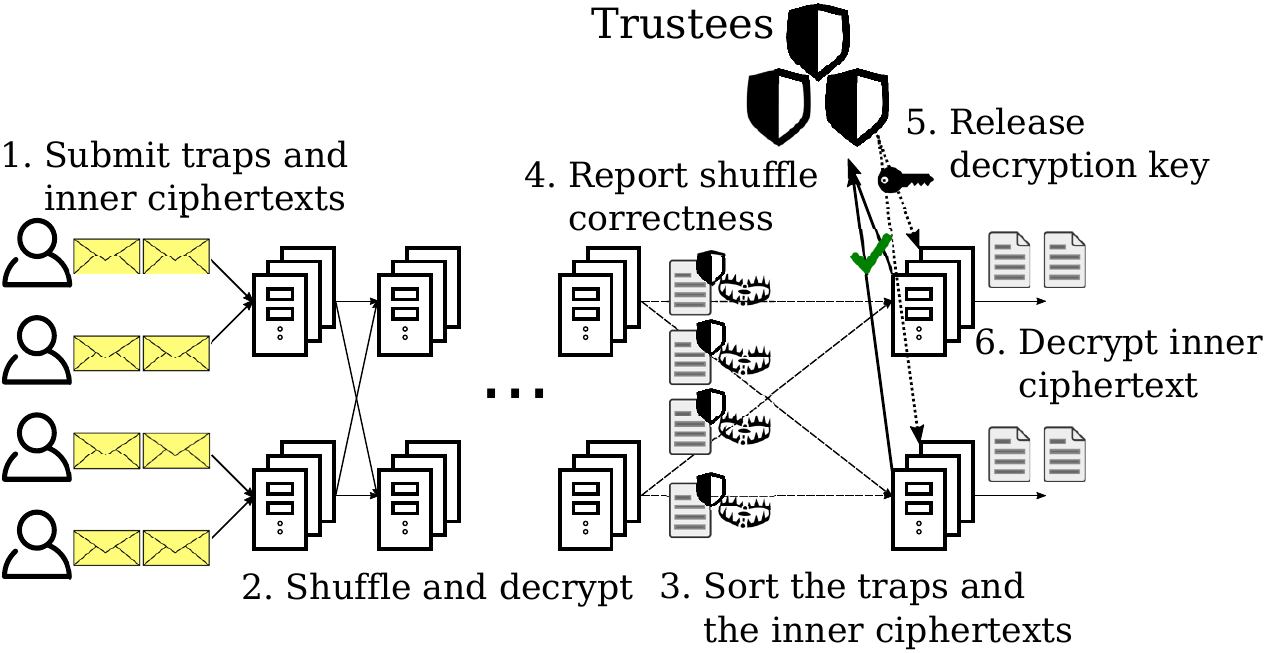}
  \caption{\sys with trap messages and trustees.}
  \label{fig:trap}
\end{figure}

Each trustee releases its share of the decryption key
if and only if every server in every group reports no violation
and the total number of traps is the same as the total number of inner ciphertexts.
Otherwise, each trustee deletes its share of the secret key.
If the trustees release the decryption key,
then each server decrypts the inner ciphertexts
to recover the actual messages.
Figure~\ref{fig:trap} summarizes the protocol.

\paragraph{Security analysis.}
The messages sent in one round cannot be replayed in another round
because the group keys change across rounds.
Thus, we only consider the security of one round.
The servers cannot tamper with any trap messages
since the honest server in each group
holds the commitments to all traps it is expecting to see.
Moreover, because the traps are mixed independently of the real messages,
the final locations of traps do not leak any information about the real messages.
We then use IND-CCA2 encryption, which creates non-malleable ciphertexts,
to prevent the adversary from tampering with the inner ciphertexts to
create related ciphertexts.
The adversary could still, however, duplicate, drop, or replace
an inner ciphertext.

The servers first explicitly check for duplicates, and abort the protocol
upon finding any.
When a malicious server removes or replaces a ciphertext,
there is at least 50\% chance that the modified ciphertext is a trap message
because the users submit the ciphertexts in a random order
and the ciphertexts are indistinguishable.
Thus, if a server drops or replaces a single ciphertext,
it causes the entire protocol run to abort with probability 50\%.
The adversary can, however, remove or replace $\kappa$ messages successfully
with probability $2^{-\kappa}$.
Since each successful tampering reduces
the anonymity set of all users by one,
the adversary can reduce the anonymity set size by
at most $\kappa$ with probability $2^{-\kappa}$.
This attack does not impact the privacy of the tampered ciphertexts:
the removed inner ciphertexts are always encrypted under
at least one honest server's key,
and thus the plaintext messages of the replaced messages are never revealed.

\subsection{Tolerating server churn} \label{sec:server_churn}
The failure of any server--even a benign one--in the protocols described thus far prevents
the failed server's group from making progress.
This is an important challenge for \sys:
with hundreds or thousands of volunteer servers involved
in each round, server failures are bound to happen.
To address this issue,
we modify \sys to use threshold anytrust groups
that we call ``many-trust groups''.
We construct the groups such that
there are at least $h$ honest servers in each group,
and enable each group to tolerate up to $h-1$ failures.

When using \sys with many-trust groups, we replace each
group public key with the key of a threshold cryptosystem.
In a $k$-server group, we share the keys in such a way that
any $k - (h - 1)$ servers can decrypt messages encrypted for the group's key.
Since there are at least $h$ honest servers in each group,
any subset of $k - (h - 1)$ servers contains an honest server.
With such groups, \sys works similar to the anytrust variant,
except that only $k - (h-1)$ group members need to participate.

Our fault-tolerance mechanism requires two changes to the network setup
described in \S\ref{sec:gen_anytrust}.
First, we need to increase the group size $k$
to ensure existence of $h$ honest servers per group.
For example, when $h = 2$, $f = 20\%$, we need $k \geq 33$
to achieve failure probability $< 2^{-64}$ (compared to $k \geq 32$ when $h=1$).
In the common case, however, only $32 = k - (h-1)$
servers need to handle the messages,
meaning the messaging latency does not increase in this case.
Appendix~\ref{app:group_size} shows how to compute $k$,
and how large $k$ must be for different values of $h$.

Second, each group must generate its threshold encryption keys.
In \sys, we use a dealer-less distributed verifiable secret sharing
(DVSS) protocol~\cite{dvss} to generate the keys
to avoid having a single trusted dealer who knows the secret key.
When a group is first formed, all servers in the group participate
in a round of the DVSS protocol.
The public output of this protocol is the group public key
and the secret key is secret shared among the $k$ servers.
For subsequent rounds, the servers can perform this operation
in the background as they mix the messages for the current round.

\sys also provides a way to recover from more than $h-1$ failures in a group
using what we call \emph{buddy groups}.
When groups are formed in the beginning of a round,
each group picks one or more buddy groups.
Each server then secret shares its share of the group private key
with the servers in each of the buddy groups.
When more than $h-1$ servers in a group fail, a new anytrust group is formed.
Each server in the new group then collects the shares of the private key
from one of the buddy groups,
and reconstructs a share of the group private key.
This allows \sys to recover from a group failure
as long as one of its buddy groups is online.
In principle, \sys could use an extra anytrust group of highly
available servers not in the actual network as a buddy group for all groups
to minimize the chance of network failure;
for example, the trustee group can be used for this purpose
in the trap-variant of \sys.

\subsection{Malicious users in \sys} \label{sec:malicious_users}
In the NIZK variant of \sys, malicious users cannot cause a protocol run to halt.
In the trap variant, however, malicious users could potentially
disrupt a round by submitting (1) missing, incorrect, or extra traps,
or (2) duplicate inner ciphertexts.
Since the servers check the traps only once the routing has completed,
\sys with traps unfortunately cannot proactively prevent such attacks.
However, \sys does provide a way to identify the malicious users
after a misbehavior is detected.
To identify disruptive users, all entry groups first reveal their private keys.
Then, all servers in each group decrypt the ciphertexts they received,
and publish the resulting traps and inner ciphertexts to all other servers,
along with the original sender of each message and the commitments of the traps.
Each server then checks if each decrypted trap matches the corresponding commitment
and vice-versa, and reports any user who fails this check.
It also reports any users who submitted the same inner ciphertexts.
Once the malicious users' identities are known, the system maintainer
could take appropriate actions (e.g., blacklist the users).
If the DoS attack is persistent after many rounds,
\sys can fall back to using NIZKs, effectively trading off
performance for availability.

\subsection{Organizing servers}
\paragraph{Ensuring maximal server utilization.}
To maximize the performance, we need
to fully utilize all servers at all times.
A na\"ive layout of the servers, however,
will cause a lot of idle time.
For example, the second server in a group cannot do any useful work until
the first server finishes.
To ensure that every server is active as much as possible,
we ``stagger'' the position of a server when it appears in different groups
(e.g., server $s$ is the first server in the first group,
second server in the second group, etc.).
This can help minimize idle time.
Changing the positions of the servers within a group does not impact
the security of the system, since the security only depends on
the existence of an honest server.

\paragraph{Pipelining.}
In scenarios where throughput is more important than latency,
\sys can be pipelined.
When we organize the servers,
we can assign different sets of servers to different layers of our network.
The network can then be pipelined layer by layer,
and output messages every one group's worth of latency.
We do not explore this trade-off in this paper,
as latency is more important for the applications we consider.

\section{Implementation and applications} \label{sec:apps}
We implemented an \sys prototype in Go
in approximately 3,500 lines of code,
using the Advanced Crypto Library~\cite{dedis-crypto}.
Our prototype implements both protection against active attacks
(\S\ref{sec:verifiable} and \S\ref{sec:trap})
and our fault-tolerance protocol (\S\ref{sec:server_churn}).
We use the NIST P-256 elliptic curve~\cite{p256} for our cryptographic group,
threshold ElGamal encryption~\cite{RoblingDenning:1982:CDS:539308} for
our fault-tolerance scheme,
and Neff's verifiable shuffle technique for
the zero-knowledge proof of shuffle correctness~\cite{neff}.
For our IND-CCA2-secure encryption scheme,
we use a key encapsulation scheme with ElGamal~\cite{cryptoeprint:2001:112},
described in Appendix~\ref{app:enc}.
The source code is available at \url{github.com/kwonalbert/atom}.
We now highlight two particularly suitable applications for \sys.

\paragraph{Microblogging.}
Microblogging is a broadcast medium in which users send short messages.
For example, Twitter is a microblogging service that supports
messages up to 140~characters. In several cases, anonymity is
a desirable property for microblogging: protest organizers can announce
their plans, and whistleblowers can publicly expose illegal activities
without fearing repercussions. \sys is a natural fit for such
applications. To blog, a user sends a short message through the \sys
network. The servers then put the plaintext messages on a public
bulletin board where other users can read them.
We use 160~byte messages in our evaluation.

\paragraph{Dialing.}
Several private communication systems~\cite{vuvuzela, alpenhorn, pung}
require a \emph{dialing} protocol by which pairs of users
can establish a shared secret.
\sys supports a dialing protocol, similar to that
of Vuvuzela~\cite{vuvuzela} and Alpenhorn~\cite{alpenhorn}.
To dial another user Bob, the initiator Alice encrypts her public key
using Bob's public key. Then, she sends her encrypted key and
Bob's identifier $id$ (e.g., his public key) through the \sys network.
The servers at the last layer put the encrypted keys into \emph{mailboxes}
based on the identifier. There are $m$ mailboxes, and each dialing
message is forwarded to mailbox $id \bmod m$.
Finally, Bob downloads the contents of the mailbox that contains the requests for him,
decrypt the messages, and establishes a shared key with Alice.

To hide the number of dialing calls a user receives,
we employ the differential privacy technique proposed by Vuvuzela.
We require one of the anytrust groups (which could be the trustees in the
trap variant) to generate dummy dialing messages for each mailbox,
where the number of dummies is determined using differential privacy.
Then, these dummy messages are distributed evenly across the Atom network,
and routed along with the actual users' dialing messages.
We refer the readers to prior work~\cite{vuvuzela} for
detailed privacy analysis of the required number of dummies.

The size of each dialing message can differ depending on the cryptosystem
used and the security guarantees.
In the simplest version in which a user simply sends an encrypted public key
of an elliptic curve cryptosystem,
the message size can be as small as 80~bytes
(Bob's public key + AEAD encrypted Alice's public key).
For more sophisticated dialing protocols, the messages can be larger.
Alpenhorn~\cite{alpenhorn}, for instance, uses identity-based
encryption, and each message is approximately 300~bytes.
Our prototype implements the simpler 80~byte message dialing scheme.

\section{Evaluation} \label{sec:eval}
To evaluate \sys, we performed two classes of experiments.
In the first set of experiments,
we measured the performance of a single anytrust group.
In the second, we perform end-to-end experiments using a large number of machines.
In these latter experiments, we verify that the system can
(1) handle a large number of messages,
and (2) scale horizontally.
We carried out our experiments in the us-east-1 region of Amazon EC2,
but we artificially introduced a latency between 40 and 160~ms
for each pair of servers (using \texttt{tc} in Linux)
to emulate a more realistic network environment.
The machines communicated over TLS channels.

\subsection{Anytrust group performance} \label{sec:single_group}
\paragraph{Experimental setup.}
To understand the performance of the building blocks of an \sys network,
we measured the latency of the cryptographic operations
and one mixing iteration.
We used Amazon EC2's c4.xlarge instances for most of
the experiments in this section,
which have four Intel Haswell Xeon E5-2666 vCPUs with 7.5 GB of memory.
For these experiments, we fix the message size at 32 bytes.
The latency increases linearly with the message size,
as we use more points to embed larger messages;
i.e., a 32-byte message is one elliptic curve point, a 64-byte message
is two elliptic curve points, etc.

\paragraph{Cryptographic primitives.}
Table~\ref{tab:crypto} shows the performance
of the cryptographic primitives used by \sys.

\begin{table}[tb]
  \caption{Performance of the cryptographic primitives. \vspace{-2mm}}
  \centering
  {\small
  \begin{tabular}{r c c}
  \hline
  \textbf{Primitives}                 & \multicolumn{2}{c}{\textbf{Latency (s)}} \\ \midrule
  $\enc$                              & \multicolumn{2}{c}{$1.40 \cdot 10^{-4}$} \\
  $\reenc$                            & \multicolumn{2}{c}{$3.35 \cdot 10^{-4}$} \\
  $\shuffle$ (1,024 messages)         & \multicolumn{2}{c}{$1.07 \cdot 10^{-1}$} \\ \hline
                                      & Prove                &  Verify             \\
  $\encprf$                           & $1.62 \cdot 10^{-4}$ & $1.39 \cdot 10^{-4}$ \\
  $\reencprf$                         & $6.55 \cdot 10^{-4}$ & $4.46 \cdot 10^{-4}$ \\
  $\shufprf$ (1,024 messages)         & $7.57 \cdot 10^{-1}$ & $1.41 \cdot 10^{0}$  \\
  \end{tabular}
  }
  \label{tab:crypto}
\end{table}

\begin{table}[tb]
  \centering
  \caption{Latency to create an anytrust group.\vspace{-2mm}}
  {\small
  \begin{tabular}{r c c c c c c}
  \hline
  \textbf{Group size}         & 4    & 8     & 16    & 32     & 64   \\
  \textbf{Setup latency} (ms) & 7.4  & 29.4  & 93.3  & 361.8  & 1432.1 \\\hline
  \end{tabular}
  }
  \label{tab:setup_latency}
\end{table}

\paragraph{Group setup time.}
Table~\ref{tab:setup_latency} shows the latency to generate an anytrust group
of different sizes.
The threshold key generation (DVSS~\cite{dvss}) is the dominating cost.
For all group sizes of fewer than 64 servers,
the setup takes less than two seconds.
In practice, we expect this overhead to be even lower,
since the servers can organize themselves into groups in the background
as they mix the messages for the current round.
\begin{figure}[tb]
  \centering
  \includegraphics[width=\linewidth]{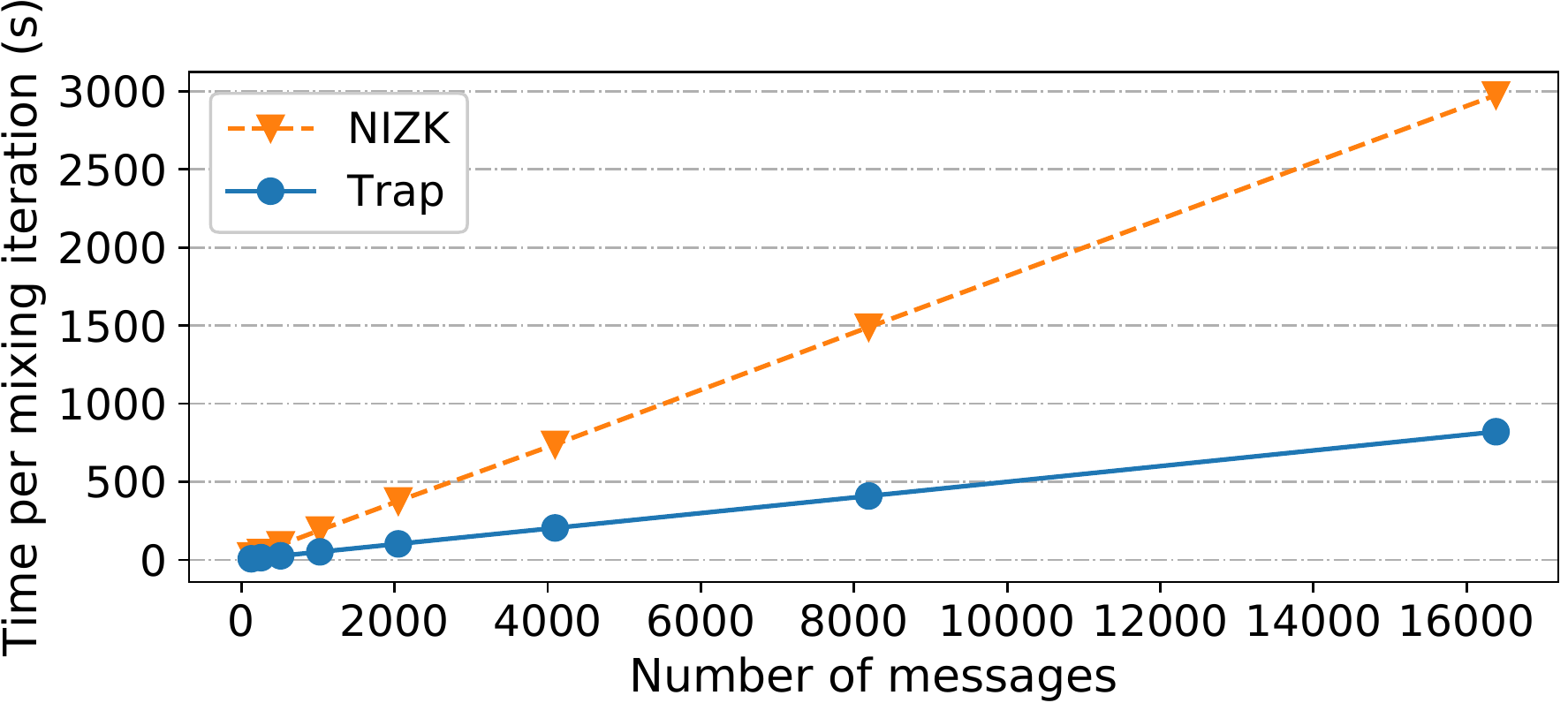}
  \caption{Time per mixing iteration for a single group
    of 32 servers as the number of messages varies.}
  \label{fig:ver_v_trap}
\end{figure}
\paragraph{NIZKs vs. traps.}
To compare the performance of the two techniques for
defending against malicious servers,
we created a single anytrust group with 32 servers,
and measured the time required to complete one mixing iteration.
The number of messages varied from 128 to 16,384 messages,
which is the expected range of message load per group.
In the trap version of \sys,
we accounted for the trap messages as well,
which doubled the actual number of messages handled by each group.
For example, if there are 1,024 groups and $2^{20}$ messages,
each group would handle 1,024 messages in the NIZK variant
and 2,048 messages in the trap variant.

As shown in Figure~\ref{fig:ver_v_trap}, the mixing time of both modes
increases linearly with the number of messages,
since the mixing time largely depends on the number of ciphertexts
each server has to shuffle and reencrypt.
The NIZK variant takes about four times longer than the
trap variant due to costly proof generation and verification.
Based on this, we estimate that a full \sys network
using NIZKs would be four times slower than a trap-based \sys network.

\begin{figure}[tb]
  \centering
  \includegraphics[width=\linewidth]{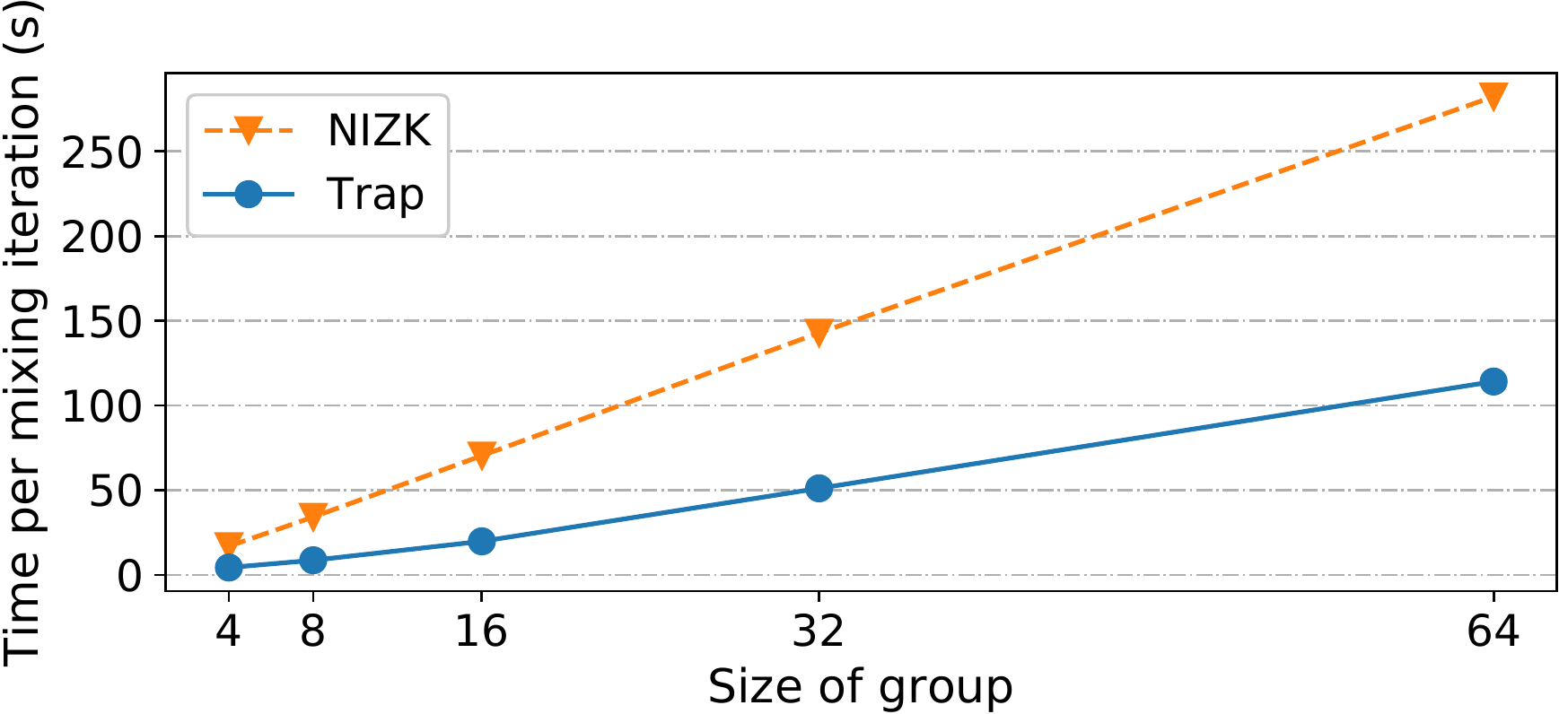}
  \caption{Time per mixing iteration for a single group
    when routing 1,024 messages as the group size varies.}
  \label{fig:group_size}
\end{figure}

\paragraph{Group size.}
The size of each anytrust group in \sys
depends on the security parameter and $f$,
the fraction of servers that are adversarial.
Figure~\ref{fig:group_size} demonstrates the impact of group size on the mixing
iteration time when the group handles 1,024 messages.
For both schemes, the mixing time increases linearly with the group size,
since each additional server adds another serial
set of shuffling and reencryption operations.
While our fault tolerance parameter $h$ impacts the group size $k$ as well,
only $k-(h-1)$ servers handle the messages in a given iteration
since $k-(h-1)$ servers is enough to decrypt the threshold encrypted ciphertexts.

\begin{figure}[tb]
  \centering
  \includegraphics[width=\linewidth]{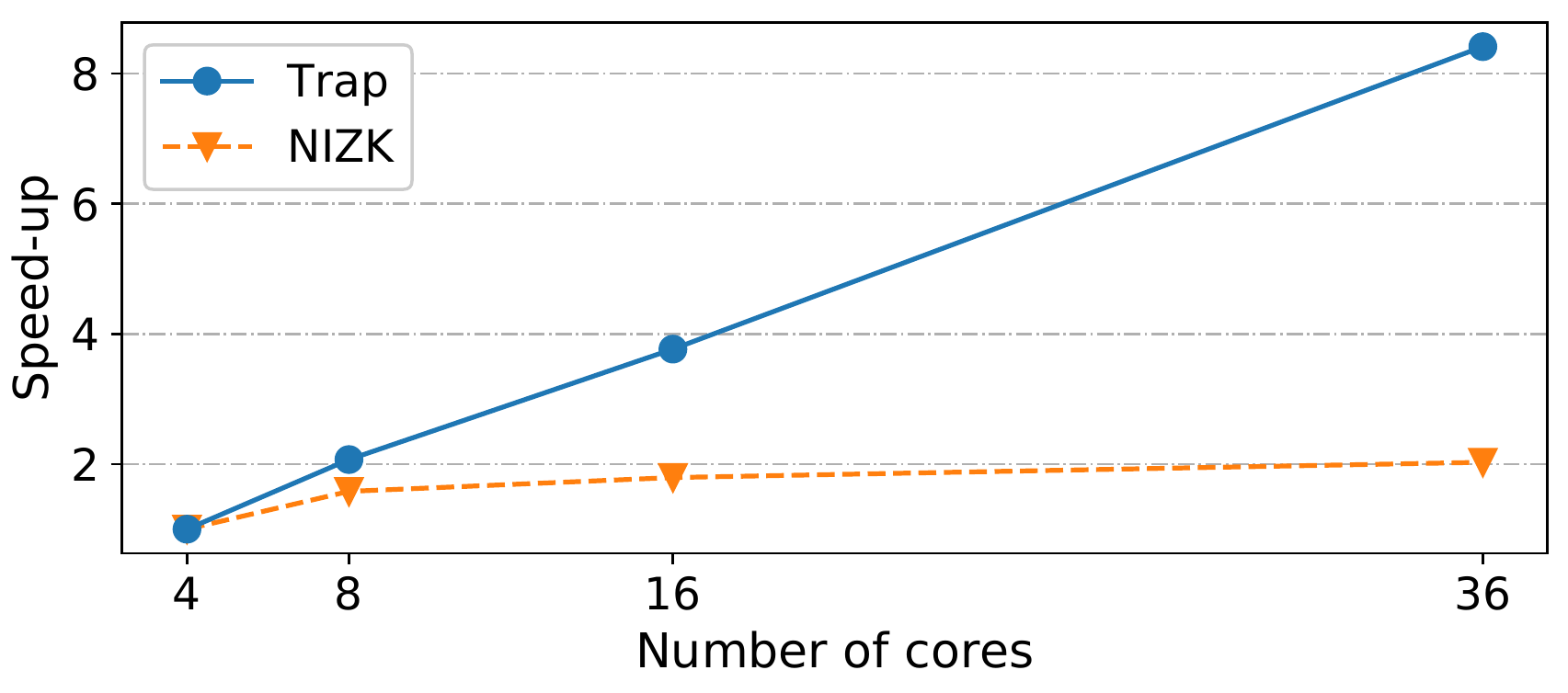}
  \caption{Speed-up of one mixing iteration of \sys as we increase
    the number of cores on a server.
    The baseline is when all servers have four cores.}
  \label{fig:latency_msgs}
\end{figure}

\paragraph{Number of cores.}
The computations that the \sys servers perform are highly parallelizable,
especially for the trap variant.
Adding more cores to each server decreases the overall latency in proportion.
To demonstrate this effect, we created an anytrust group of 32 servers
using EC2 c4 nodes with 4, 8, 16, and 36 cores,
and we routed 1,024 messages through them.
Figure~\ref{fig:latency_msgs} shows the speed-up of different anytrust
groups over the one consisting of only four core servers.
The speed-up is nearly linear for the trap-variant,
since majority of the load is parallelizable.
The speed-up of the NIZK variant is sub-linear
because the NIZK proof generation and verification technique
we use is inherently sequential.

\subsection{Large-scale evaluation of \sys} \label{sec:large_scale}
\begin{figure}[tb]
  \centering
  \includegraphics[width=0.45\linewidth,origin=c,angle=180]{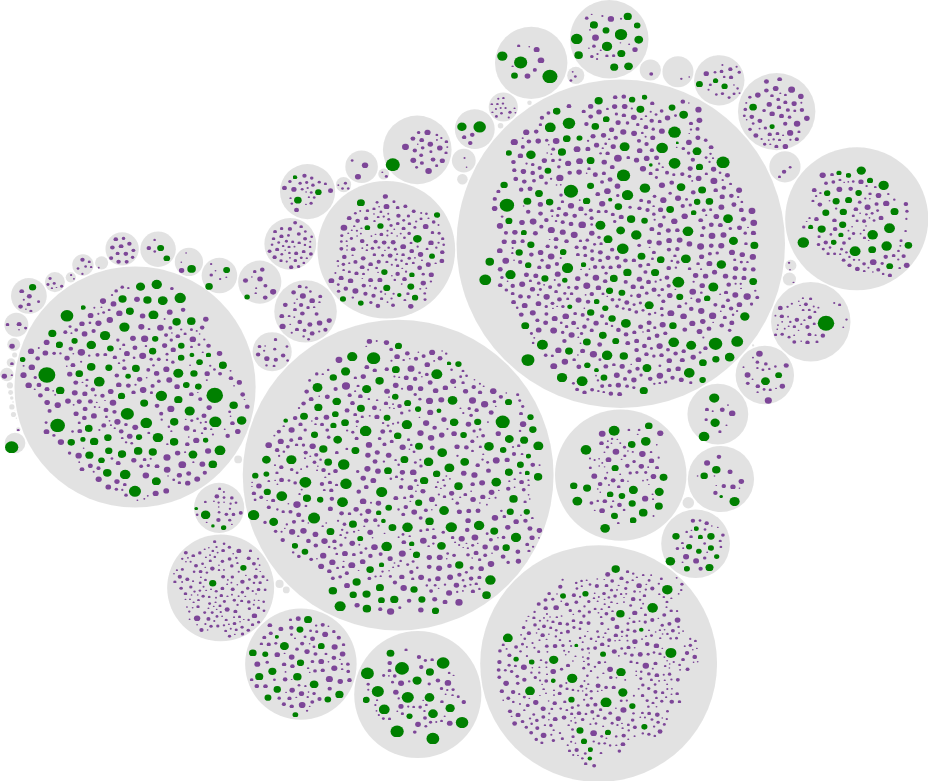}~~\rulesep{}~~%
  \includegraphics[width=0.45\linewidth,origin=c,angle=0]{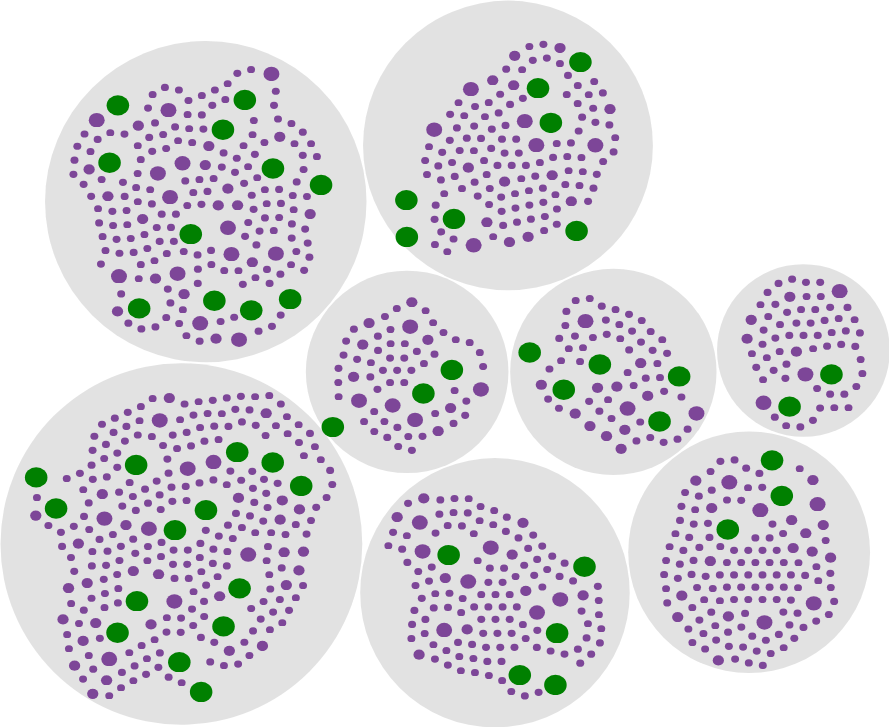}
  \caption{The Tor network topology~\cite{torfig} (left) and our network topology (right).
          The size of a node indicates its capacity; very high-capacity nodes
          are in green.
          In our topology, links within a cluster have 40~ms latency and across
          clusters have  80-160~ms latency.
          }
  \label{fig:topology}
\end{figure}

\paragraph{Experimental setup.}
In this set of experiments, we used up to 1,024 various Amazon EC2 machines
to test \sys's scalability and performance
using the trap variant of \sys.
In each experiment, 80\% of the
servers had 4 cores (c4.xlarge), 10\% had 8 cores (c4.2xlarge),
5\% had 16 cores (c4.4xlarge), and 5\% had 32 cores (c4.8xlarge).
We used the bandwidth statistics of the Tor network~\cite{tor_metric} as a proxy
to chose the servers for our network\footnote{Statistics on core counts of
  Tor servers were not available.}:
80\% of the servers have less than 100~Mbps,
10\% have between 100~Mbps and 200~Mbps,
5\% have between 200~Mbps and 300~Mbps,
and 5\% have over 300~Mbps of available bandwidth.
Figure~\ref{fig:topology} shows our network topology.

We picked the system parameters
assuming $f = 20\%$ of the servers are malicious.
We set up our groups to handle
one server failure for each group (\S\ref{sec:server_churn}).
Thus, we set the group size to 33 servers,
and required 32 out of the 33 servers to route the messages.
Finally, we used $\mu = 13,000$, where $\mu$ is
the average number of dummy messages
by each server in an anytrust group
for differential privacy~\cite{vuvuzela}.
Thus, on average, we expect about $32 \cdot \mu = 410,000$ dummy
messages total in the network for anytrust groups of 32 servers.

We used $T=10$ iterations with the square network for the evaluation.
To measure the latency of one round of protocol run,
we measure the time lapse between
the moment that the first server in the first layer receives a message
and the last server in the last layer outputs a message.

We answer the following questions in this section:
\begin{itemize}
  \item Can \sys support a large number of users?
  \item How does \sys scale horizontally?
  \item How does \sys compare to prior systems?
\end{itemize}

\begin{figure}[tb]
    \centering
    \includegraphics[width=\linewidth]{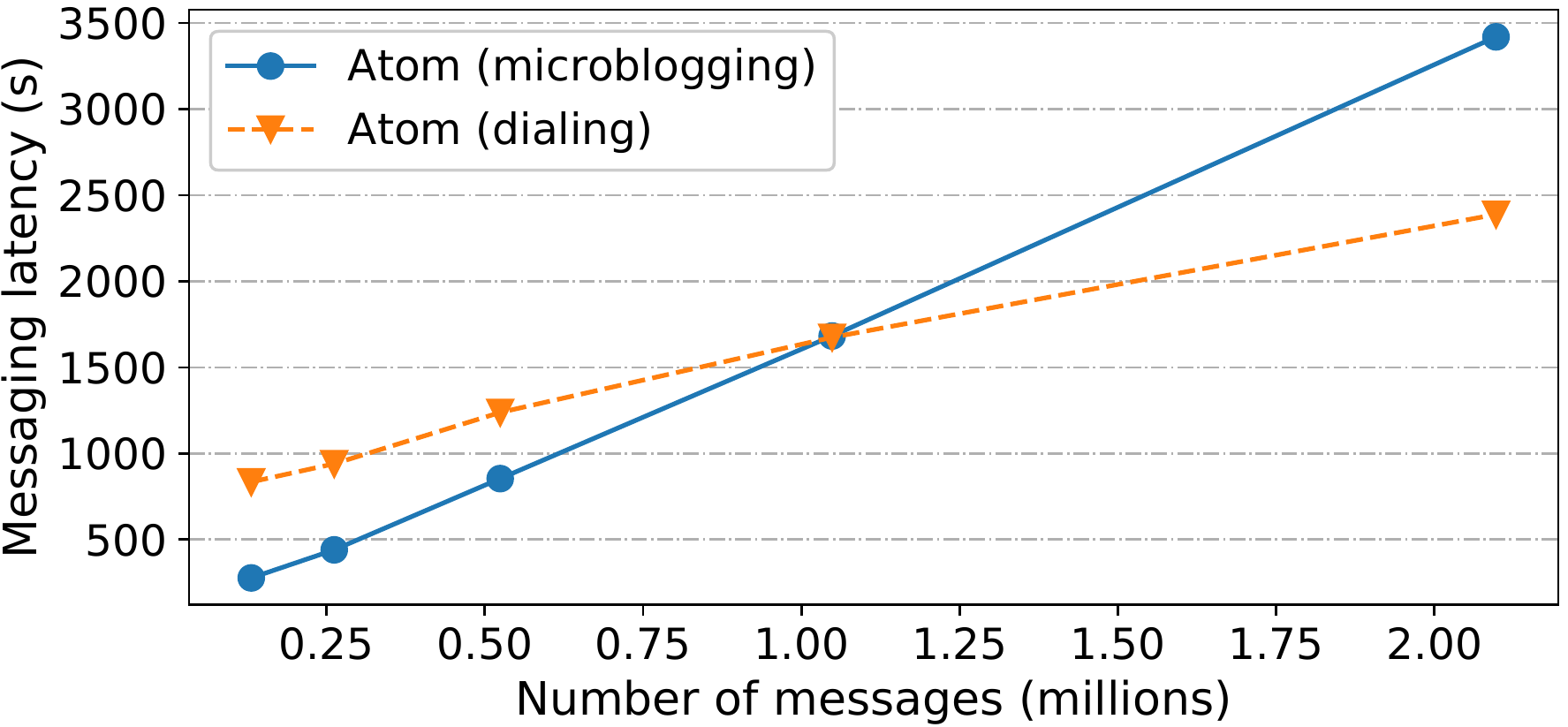}
    \caption{Latency of \sys for microblogging and dialing for varying number of messages.
      The latency increases linearly with the number of messages.}
    \label{fig:msg_v_latency_full}
\end{figure}

\paragraph{Number of messages.}
We used 1,024 servers organized into 1,024 groups, and measured the latency as
the total number of messages varied.
As Figure~\ref{fig:msg_v_latency_full} shows, the latency
increases linearly with the total number of messages,
since the number of messages handled by each group increases linearly
with the total number of messages.
The difference in the slope between the two applications
is due to the smaller message size for dialing.
For both applications,
our prototype can handle over a million users with a latency of 28 minutes.

\begin{figure}[tb]
  \centering
  \includegraphics[width=\linewidth]{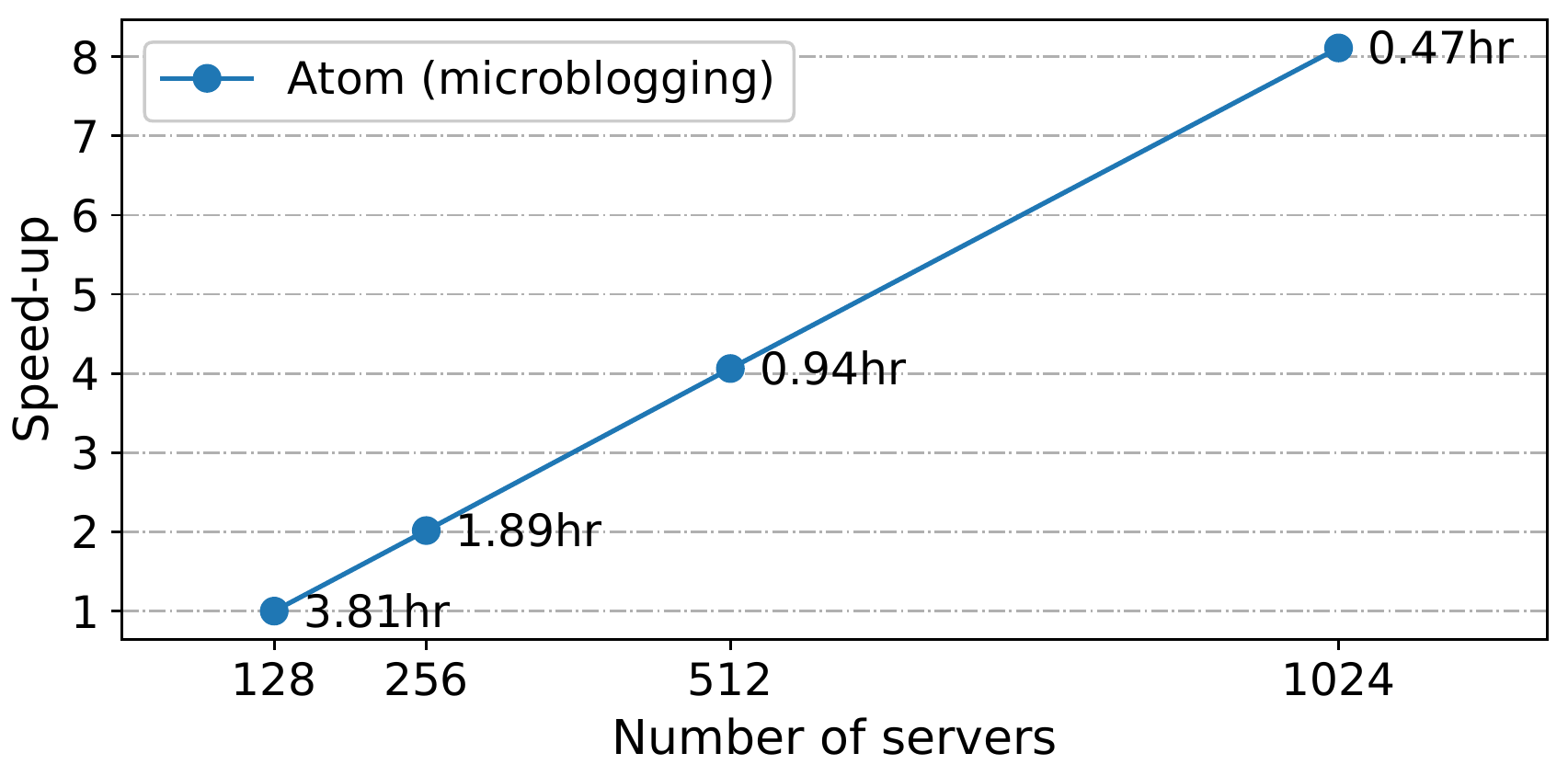}
  \caption{Speed-up of \sys networks of varying sizes relative to
    an \sys network of 128 servers.
    The speed-up is linear in the number of servers in the network.}
  \label{fig:hoziontal_scale}
\end{figure}

\begin{figure}[t]
    \centering
    \includegraphics[width=\linewidth]{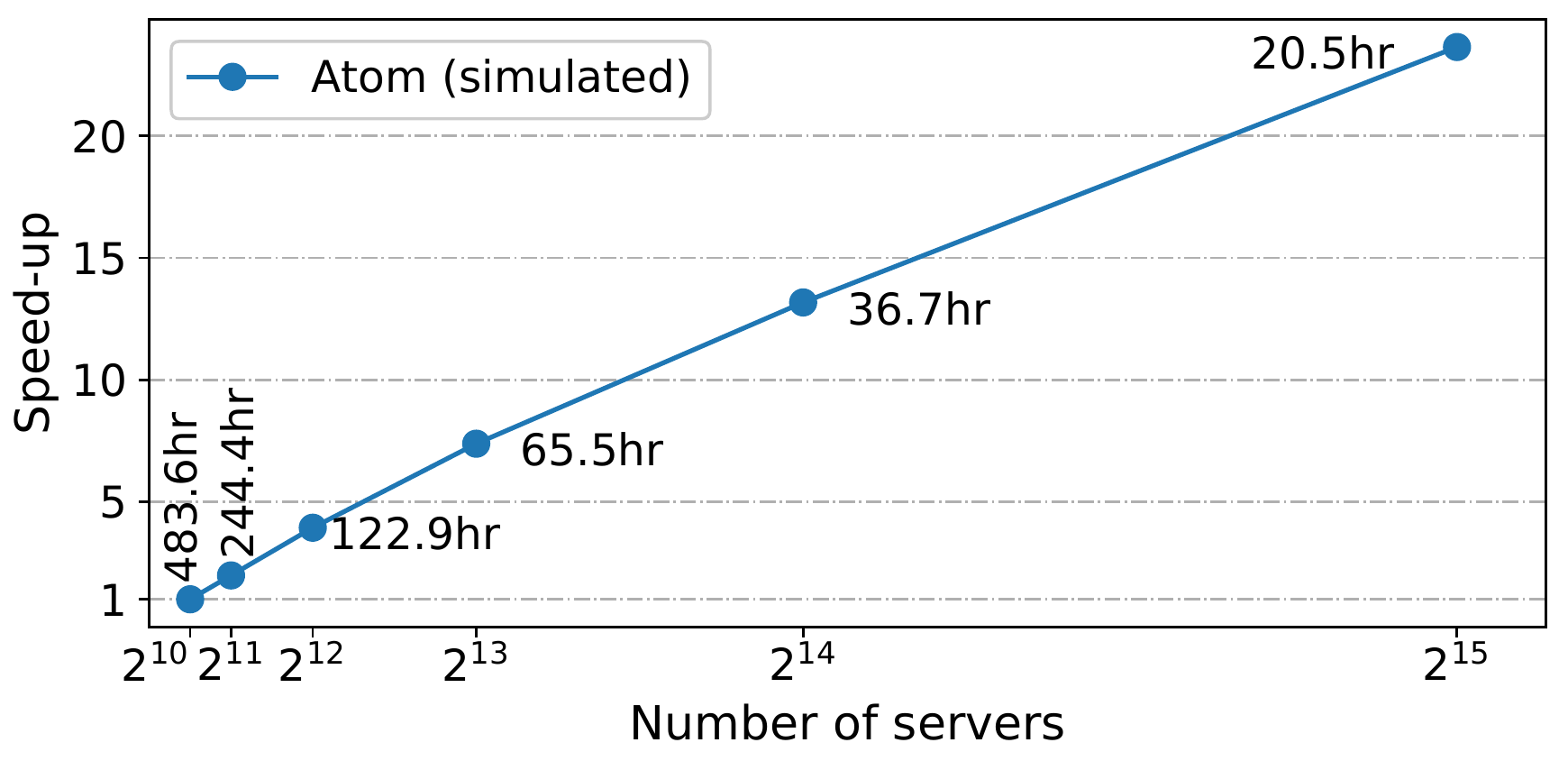}
    \caption{Simulated speed-up of \sys networks of varying sizes
      relative to an \sys network of 1,024 servers when routing
      a billion microblogging messages.
      At this scale, the speed-up is sub-linear in the number of servers.}
    \label{fig:further_scale}
\end{figure}

\paragraph{Horizontal scalability.}
To demonstrate that \sys scales horizontally, we measured
the end-to-end latency for the network to route a million
microblogging messages as the number of servers varied.
As shown in Figure~\ref{fig:hoziontal_scale},
the network speeds up linearly with the number of servers.
That is, an \sys network with 1,024 servers is twice as fast as
one with 512 servers.

We also simulated larger \sys networks to show further scalability.
Here, we modified the implementation to model the expected latency given
an input using values shown in Table~\ref{tab:crypto},
instead of actually performing the operations.
We then had each physical server in our network emulate a large
number of logical servers.
Figure~\ref{fig:further_scale} shows the simulated latency
when routing over a billion microblogging messages using larger \sys networks.
Unlike the cases with less than 1,024 servers,
we observed that the speed-up is slightly sub-linear in the number of servers.
We believe there were two reasons for this.
First, there are $G^2$ connections between two layers
where $G$ is the number of servers per layer.
When there are tens of thousands of groups per layer,
the number of connections became unmanageable for some servers.
Second, the single trustee group experienced
performance degradation.
While the trustees could handle tens of thousands of TLS connections,
the overhead of establishing TLS connection became non-negligible
at this scale.

\begin{table}[t]
  \centering
  \captionof{table}{
    Latency to support a million users.
    For microblogging, \sys is $23.7\times$ faster than Riposte
    with 1,024 servers.
    Vuvuzela is $56\times$ faster than \sys for dialing.\vspace{-2mm}
    }
  \label{tab:prior_comparison}
  {\footnotesize
  \begin{tabular}{rr|rr}
    & & \multicolumn{2}{c}{Latency (min.)} \\
    &  & \textit{Microblog} & \textit{Dial} \\
    & \multicolumn{1}{c|}{Hardware} &
    \multirow{2}{*}{$\begin{pmatrix} \textit{Speedup} \\ \textit{vs.~Riposte} \end{pmatrix}$} &
    \multirow{2}{*}{$\begin{pmatrix} \textit{Slowdown} \\ \textit{vs.~Vuvuzela} \end{pmatrix}$} \\
    & \multicolumn{1}{c|}{Config.}  & & \\ \midrule
  Atom & $128\times$mixed  & 228.7 (2.9$\times$) & 225.1 (450$\times$) \\
       & $256\times$mixed  & 113.4 (5.9$\times$) & 112.6 (225$\times$) \\
       & $512\times$mixed  & 56.3 (11.9$\times$) & 55.5 (111$\times$)  \\
       & $1024\times$mixed & 28.2 (23.7$\times$) & 27.9 (56$\times$)   \\ \midrule
  Alpenhorn~\cite{alpenhorn}& $3\times$c4.8xlarge & -- & 0.5 (1$\times$) \\
  Vuvuzela~\cite{vuvuzela} & $3\times$c4.8xlarge & -- & 0.5 (1$\times$) \\
  Riposte~\cite{riposte} & $3\times$c4.8xlarge & 669.2 (1$\times$) & -- \\ \midrule
  \end{tabular}
  }
\end{table}

\paragraph{Comparison to prior work.}
Table~\ref{tab:prior_comparison} compares the performance of \sys
to that of three prior works: Riposte~\cite{riposte},
Vuvuzela~\cite{vuvuzela}, and Alpenhorn~\cite{alpenhorn}.
Riposte is an anonymous microblogging system that
uses centralized anytrust servers.
To compare Atom's microblogging capabilities,
we used the fastest variant of Riposte which uses three 36-core machines
configured to handle a million messages.
As shown in Table~\ref{tab:prior_comparison}, Riposte takes approximately
11 hours to anonymize a million messages.
\sys can support a million microblogging messages
in 3.8 hours with 128 servers and 28 minutes with 1,024 servers,
which is $2.9\times$ and $23.7\times$ faster than Riposte.
While \sys uses many more servers to achieve better performance,
Riposte cannot take advantage of more servers without additional
security assumptions.
In particular, Riposte could scale by replacing
each logical server with a cluster of physical servers.
An adversary, however, still needs to only compromise one server
from each cluster to break the security of the system.

Vuvuzela and Alpenhorn are private messaging systems
with centralized anytrust servers that also support dialing.
We show the dialing latency with one million users
for the two systems when the network
consists of three 36-core machines with 10~Gbps connection between them.
These two systems are approximately $56\times$ faster
than \sys for mainly two reasons: (1) more efficient cryptography,
as they use hybrid encryption, while \sys uses public key encryption,
and (2) their machines are more powerful than the average machine in our network.
Although \sys is slower, we still believe it is suitable for dialing.
For example, Alpenhorn~\cite{alpenhorn} suggests running a dialing round
once every few hours due to client bandwidth limitations.
\sys can support more than two million users on this time scale.

\sys also offers three concrete benefits over Vuvuzela and Alpenhorn.
First, the bandwidth requirement for each server is significantly lower.
Vuvuzela servers use 166~MB/sec of bandwidth for all servers~\cite{vuvuzela},
while \sys servers use less than 1~MB/sec of bandwidth.
Second, \sys provides a clear way to scale further:
adding more servers to our network will reduce the overall latency.
Vuvuzela or Alpenhorn could replace each logical server
with a cluster of servers to scale-out.
However, similar to Riposte, the adversary only needs to compromise
a server in each cluster to break the security of the system.
Finally, \sys can provide additional privacy
by preventing servers from tampering with honest users' messages.
Vuvuzela and Alpenhorn do not aim to prevent servers from tampering
with the users' messages, and as a result, a malicious server
can drop all but one honest user's messages
(the honest user will still be protected via differential privacy).
In \sys, we guarantee that the users will enjoy anonymity among all honest users
in addition to differential privacy.

\section{Discussion and future work} \label{sec:discussion}
We now discuss some aspects of \sys we did not consider
in this paper and describe potential future work.

\paragraph{Estimated deployment costs.}
The deployment cost of a server depends on compute and bandwidth cost.
When renting compute power from a commercial cloud service,
the cost is usually fixed per hour of uptime.
For example, if a volunteer wants to maintain a 100\% up-time
with a four core server on Amazon AWS, it would cost about \$146/month,
while a 36-core server would cost about \$1,165/month
as of September 2017.

The bandwidth cost depends on the number and the size of messages
routed by the server, but we can estimate an upper bound
by calculating the bandwidth requirement to rate-match the compute power of the server.
Based on our microbenchmarks of the cryptographic primitives
(Table~\ref{tab:crypto}),
a four core server in the trap variant of \sys
can reencrypt about 2,700 messages/second
and shuffle about 9,200 messages/second when each message is 32 bytes.
This translates to about 90KB/second and 300KB/second of bandwidth respectively.
Assuming a constant flow of 300KB/second,
the upper bound on the bandwidth cost would be about
\$7.20/month on AWS.
For a 36-core server, this cost would scale linearly,
and would cost about \$65/month.

\paragraph{Denial-of-service.}
Our focus in this work was on providing scalable anonymity
in the face of a near-omnipresent adversary.
We explicitly left availability attacks out of scope.
Our fault-tolerance mechanism (\S\ref{sec:server_churn}), however,
can help defend against a small number of malicious failures as well:
as long as there are fewer than $h$ failures per group, benign or malicious,
\sys can recover.
Defending against a large scale DoS (e.g., in which the attacker takes
more than half of the servers offline) remains an important challenge
for future work.
Proactively defending against malicious users in the trap variant
is another important future work,
as \sys can only retroactively identify malicious users
after a disruption happens (\S\ref{sec:malicious_users}).

\paragraph{Load balancing.}
In a real-world \sys deployment, there will be some variances in the capacity
of the servers;
e.g., number of cores, amount of available bandwidth, memory, etc.
From a performance perspective, it would be beneficial to
have the more powerful servers appear in more groups.
Such non-uniform assignments of servers to groups, however, could
result in an adversary controlling a full \sys group.
We must therefore be careful when load-balancing servers,
but the degradation in security may be worth the performance gain in practice.
Tor~\cite{tor}, the only widely deployed anonymity network,
makes this trade-off:
servers with more available bandwidth are more likely picked
when forming a Tor circuit.

\paragraph{Intersection attacks by servers.}
Apart from the intersection attacks~\cite{kedogan2002limits, danezis2004statistical}
naturally occurring due to the changes
in the set of \sys users,
malicious servers in the trap variant of \sys could attempt to
launch an intersection attack by selectively removing a user's ciphertexts.
It is not possible, however,
to remove both trap and message ciphertexts without getting caught,
since the commitments of the traps are made available to all servers.
Thus, the malicious server has to guess the real ciphertext to remove.
As a result, the adversary performing such an attack on one user is caught
with probability 50\%, and this probability is amplified to $2^{-\kappa}$
for $\kappa$ trials. Therefore, while \sys does not completely prevent
intersection attacks by the servers,
it does limit the number of times the adversary can attempt
this attack.

\section{Related work}\label{sec:related}
Tor~\cite{tor} is the only anonymous communication system
in widespread use today.
Like \sys, Tor scales horizontally.
Unlike \sys, Tor aims to support low-latency real-time traffic streams,
and Tor does not aim to defend against a global network adversary.
Recent analysis of the Tor network suggests that even certain local adversaries
may be able to de-anonymize Tor
users~\cite{Herrmann, Panchenko, ccs2012-fingerprinting, wang14-fingerprinting-defenses}.

Free-route mix-nets~\cite{aqua,crowds,mixmaster,mixminion}
also scale horizontally.
As with Tor, these systems do not provide strong anonymity
properties in the face of powerful global adversaries.
When the adversary can monitor the entire network and control
some servers, the anonymity properties of these systems can degenerate
to the set of users who share the same entry point to the network.

Mix-nets~\cite{mixnet} and
Dining Cryptographer networks (DC-Nets)~\cite{dcnet} are the earliest examples
of anonymity systems that provide protection against global adversaries.
However, neither of these systems scales horizontally:
mix-nets incur overhead linear in the number of servers,
and DC-Nets incur overhead quadratic in the number of participants.
Systems that build on these primitives~\cite{herbivore,dissentv2,riffle}
face similar problems when trying to scale.
Riposte~\cite{riposte} is an anonymous microblogging system that uses techniques from private information
retrieval~\cite{chor-pir} to scale to millions of users.
Like a DC-Net, Riposte requires each server to perform work quadratic in the number of messages
sent through the system.

The parallel mix-net of Golle and Juels~\cite{golle2004parallel} uses
a distributed network of mix servers, similar to \sys.
In this mix-net, Borisov showed that if the adversary controls some
inputs and learns the output positions of the controlled inputs,
then the adversary can infer some information about the messages
sent by the honest users~\cite{Borisov2006}.
In contrast, knowing output positions of some users' messages in \sys
does not result in anonymity loss for the other users.

Vuvuzela~\cite{vuvuzela} and Alpenhorn~\cite{alpenhorn}
are two recent private-messaging systems
that also provide \emph{dialing} mechanisms that
a user can use to establish a shared secret with another user.
Both systems require all messages to pass through
a centralized set of servers, making the systems scale only vertically.
Moreover, contrary to \sys,
malicious servers in these systems can drop all but one honest
user's messages. In \sys, we
ensure no honest users' messages are tampered with.
Thus, \sys can provide anonymity in addition to differential privacy.

Vuvuzela also provides point-to-point
metadata-hiding communication (not anonymity).
Using Vuvuzela, two users who share a secret can communicate via the system
without an adversary learning that these two users are communicating.
Vuvuzela, however, cannot be used to anonymously organize protests,
since the recipient of a message always knows its sender.
Pung~\cite{pung} addresses the same problem as Vuvuzela, but does so
without the need for an anytrust assumption.
Pung instead relies on computational private information retrieval, which escapes the
need for trust assumptions but comes with significant computational costs~\cite{kushilevitz2000one}.

Stadium~\cite{stadium} is a recent work that aims to horizontally
scale Vuvuzela (i.e., private point-to-point communication)
using distributed anytrust groups in a similar way to \sys.
Stadium uses a system architecture inspired by the parallel
mix-net~\cite{golle2004parallel} instead of permutation networks,
and uses verifiable shuffle and cover traffic (dummy messages)
to achieve a differential private notion of security.
Stadium achieves lower latency than \sys by
verifiable shuffling only the metadata of each message (e.g., a digest).
The actual messages are encrypted using efficient hybrid encryption,
and the servers after each iteration of mixing
check that the metadata matches the ciphertexts.
This strategy cannot be used in \sys since a user does not know
the path that her message will take.
Instead, Atom provides anonymous broadcasting without cover traffic
and security based on indistinguishability of permutations
at a higher latency.

Loopix~\cite{loopix} is a recent system that provides asynchronous
bidirectional anonymous messaging, and scales horizontally.
Loopix can provide low-latency communication
using servers that insert small amount of delays before routing
the messages.
However, the security guarantees of Loopix degrade as the fraction of
adversarial servers increases.
In contrast, \sys provides a way
to remain secure even when a large fraction of servers are actively
malicious, at the cost of polylogarithmic increase in the latency due
to larger group sizes. MCMix~\cite{mcmix} is another system that
provides bidirectional anonymous messaging, but using multiparty
computation (MPC). MCMix achieves better performance
than \sys by defending against a weaker adversary.
The MCMix prototype, for example, only supports
three-server MPC that provides security against one passively
malicious server.

The security analysis of \sys draws on the theoretical analysis of permutation networks.
Permutation networks have long been studied as a way to permute
a large number of elements using a network built of small
components (``switches'')~\cite{waksman1968permutation}.
In a 1993 paper, Rackoff and Simon~\cite{rackoff1993cryptographic} proposed building
a distributed mix-net from a large permutation network.
Their scheme required $O(\log^k n)$ iterations of mixing to mix $n$
messages, where $k$ was a ``double-digit'' number~\cite{czumaj1999delayed}.
In \sys, we convert this theoretical result into a practical one.
Abe also proposed building a distributed mix-net from
a butterfly permutation network~\cite{abe1999mix}, though
the construction had a flawed security analysis~\cite{abe2001remarks}.

Recent theoretical work investigated
the number of iterations of a butterfly network required
so that a random setting of the switches produces a random
permutation~\cite{morris2008mixing,morris2009improved,morris2013improved}.
Czumaj and V\"ocking~\cite{czumaj2014} have recently argued
that $O(\log^2(M))$ iterations are enough to generate an ``almost'' random permutation of $M$ inputs.
Czumaj~\cite{czumaj2015} also studied
randomly constructed networks
and demonstrated that most networks of depth $O(\log^2(M))$ and width $O(M)$
produce good random permutations.
H{\aa}stad studied a permutation network~\cite{haastad2006square} based on
shuffling the rows and columns of a square matrix.
Any of these networks could be used as the underlying topology for an \sys network,
but we focused on using the network by H{\aa}stad~\cite{haastad2006square}
due to its efficiency.

\section{Conclusion}\label{sec:conc}
\sys is a traffic-analysis resistant anonymous messaging system
that scales horizontally.
To achieve strong anonymity and scalability,
\sys divides servers into many anytrust groups,
and organizes them into a carefully constructed topology.
Using these groups, we design an efficient protocol for collectively
shuffling and rerandomizing ciphertexts to protect users' privacy.
We then propose two mechanisms,
based on zero-knowledge proof techniques and trap messages,
to protect against actively malicious servers.
Finally, we provide a low-overhead fault-recovery mechanism for \sys.
Our evaluation of \sys prototype on a distributed
network consisting of over one thousand servers demonstrates that the
system scales linearly as the number of participating servers increases.
We also demonstrated that \sys can support more than a million users
for microblogging and dialing.
With its distributed and scalable design,
\sys takes traffic-analysis-resistant anonymity one step closer to real-world practicality.

\section*{Acknowledgements}
We thank Nirvan Tyagi, David Lazar, Riad Wahby, Ling Ren,
and Dan Boneh for valuable feedback
and discussion during this project.
We also thank the anonymous reviewers, and our shepherd Peter Druschel.
This work was supported in part by the NDSEG fellowship.

\appendix

\section{Cryptographic details} \label{app:enc}
We describe a modification to
the ElGamal cryptosystem~\cite{elgamal1984public} for \sys.
We work in a cyclic group $\G$ of order $q$ with generator $g$ in which the
Decision Diffie-Hellman problem is hard~\cite{boneh1998decision}.
The space of messages, ciphertexts, and public keys is $\G \cup \{\bot\}$,
where $\bot$ represents a special null element.

\begin{itemize}
  \item $(x, X) \gets \keygen()$.
    Sample $x \gets_R \Z_q$ and set $X = g^x \in \G$.

  \item $(R, c, Y) \gets \enc(X, \msg)$.
    Sample $r \gets_R \Z_q$ and set $R \gets g^r$.
    Set $c \gets \msg \cdot X^r$, and $Y = \bot$.
    $(R, c)$ forms the ElGamal ciphertext,
    and $Y$ is a new element for \sys.

  \item $\msg \gets \dec(x, (R, c, Y))$.
    Return $\msg \gets c/R^x$.
    The symbol ``$/$'' indicates multiplication by
    the inverse of the second operand.
    If $Y \neq \bot$, then the algorithm fails.

  \item $\ciphertexts' \gets \shuffle(X, C)$.
    To rerandomize a single ciphertext $(R, c, \bot)$ for public key $X$, sample
    $r' \gets \Z_q$ and compute $(g^{r'} \cdot R, c \cdot X^{r'}, \bot)$.
    If $Y \neq \bot$, then the algorithm fails.
    To shuffle, rerandomize all ciphertexts and then permute them.

  \item $(R', c', Y') \gets \reenc(x, X', (R, c, Y))$.
    If $Y = \bot$, then set $Y = R$ and $R = 1_G$,
    where $1_G$ is the multiplicative identity of $G$.
    The algorithm proceeds in two steps:
    First, remove a layer of encryption using $x$:
    $c_{\textsf{tmp}} \gets c / Y^x$.
    Then, reencrypt for $X'$:
    sample $r' \gets \Z_q$ and set $R' \gets g^{r'} \cdot R$ and $Y' = Y$.
    Set $c' \gets c_{\textsf{tmp}} \cdot {X'}^{r'}$.

    Intuitively, $Y$ holds the randomness used to
    encrypt for the current group, while $R$ holds the
    randomness used to encrypt for the next group.
    By keeping both $Y$ and $R$,
    the servers in the current group can decrypt out-of-order.
    In \sys, before the last server of a group forwards $(R', c', Y')$
    to the next group, it sets $Y' = \bot$;
    at this point, all layers of encryption by the current
    group have been peeled off, and $c'$ is encrypted only under $X'$, making
    $Y'$ unnecessary.
\end{itemize}

We use a key encapsulation scheme with ElGamal
for our IND-CCA2 encryption scheme for inner ciphertexts.
\begin{itemize}
  \item $(x, X) \gets \keygen()$.
    Sample $x \gets_R \Z_q$ and set $X = g^x \in \G$.

  \item $(R, c) \gets \enc(X, \msg)$.
    Sample $r \gets_R \Z_q$, and set $R \gets g^r$.
    Generate a shared secret $k = X^r = g^{xr}$,
    and set $c \gets \textsf{AEnc}(k, \msg)$ where $\textsf{AEnc}$ is
    an authenticated symmetric encryption scheme.
    For \sys, we used NaCl~\cite{nacl} for $\textsf{AEnc}$.

  \item $\msg \gets \dec(x, (R, c))$.
    Generate the shared secret $k = R^x = g^{xr}$.
    Set $\msg \gets \textsf{ADec}(k, c)$, where \textsf{ADec}
    is the decryption routine for $\textsf{AEnc}$.
\end{itemize}

Finally, we describe the NIZKs we use.
\begin{itemize}
  \item $(\ciphertext, \prf) \gets \encprf(\pubkey, \msg)$.
    Compute the ciphertext $(g^r, m \cdot X^r, \bot) \gets \enc(X, \msg)$,
    and keep $r$. Pick a random $s \in \Z_q$.
    Then, compute $t = H(c \| g^s \|  X)$, and $u = s + t \cdot r$,
    where $H$ is a cryptographic hash function.
    Set $\ciphertext = (g^r, m \cdot X^r, \bot)$ and $\prf = (g^s, u)$.

    To verify this proof, a verifier checks that $g^u = g^s \cdot g^{rt}$.

  \item $(\ciphertext, \prf) \gets \reencprf(\seckey, \pubkey, \msg)$.
    We use the Chaum-Pedersen proof~\cite{chaum-pedersen} without any modifications.

  \item $(\ciphertexts', \prf) \gets \shufprf(\pubkey, \ciphertexts)$.
    We use the Neff verifiable shuffle~\cite{neff} without any modifications.
\end{itemize}
For $\encprf$, the same proof $\pi$ cannot be used
for two different public keys,
since the public key $X$ is given as input to
$H$ when computing $t$.
This prevents an adversary from copying the ciphertext and the NIZK
submitted to one group, and submitting them to a different group.

\begin{figure}[tb]
  \centering
  \includegraphics[width=\linewidth]{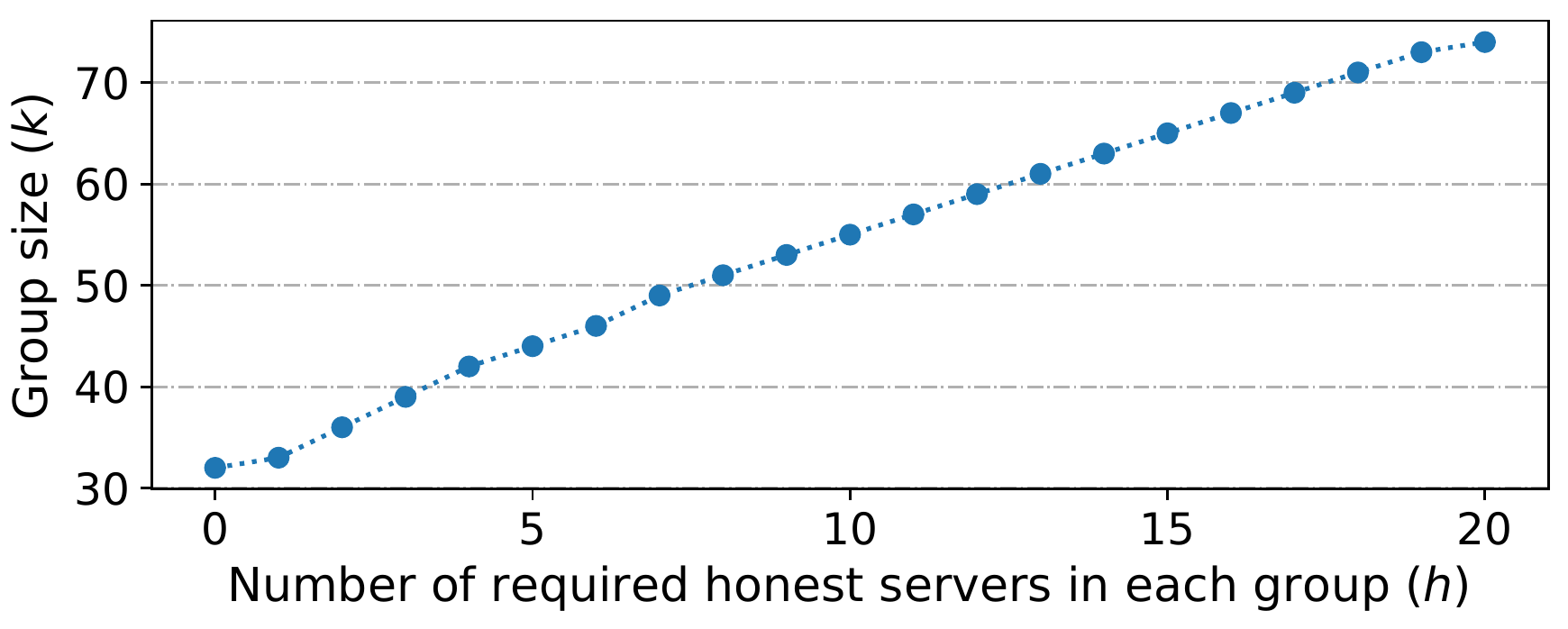}
  \caption{Required size of each group to maintain the security guarantees
    for different values of $h$ for $f=0.2$ and $G=1,024$.}
  \label{fig:h_group_size}
\end{figure}

\section{Many-trust group size} \label{app:group_size}
\sys handles server churns using many-trust groups,
as described in \S\ref{sec:server_churn}.
To tolerate up to $h-1$ server failures in a group,
each many-trust group needs at least $h$ honest servers.
We create such groups with high probability by increasing the group size.
Let $k$ be the group size, $f$ be the fraction of malicious servers,
and $G$ be the number of groups in the network.
Similar to the analysis done in \S\ref{sec:gen_anytrust} for the anytrust groups,
we need
\begin{align*}
  & G \cdot \Pr[\text{fewer than $h$ honest servers in a group of $k$ servers}] \\
  & = G \cdot \sum_{i=0}^{h-1}\Pr[\text{$i$ honest servers in a group of $k$ servers}] \\
  & = G \cdot \sum_{i=0}^{h-1}\binom{k}{i} (1-f)^if^{k-i} < 2^{-64}
\end{align*}
Figure~\ref{fig:h_group_size} shows the required size of the groups $k$
as a function of $h$ when $f=0.2$, and $G=1024$.

\newpage
\bibliographystyle{plain}
\bibliography{refs}

\end{document}